\documentclass[dvips,12pt]{article}

\usepackage{color}

\usepackage[pdftex]{graphicx}
\usepackage{url}

\usepackage[margin=0.75in]{geometry}
\usepackage[utf8]{inputenc}
\usepackage{amsmath,verbatim}
\usepackage{amsfonts}
\usepackage{amssymb}
\usepackage{adjustbox}
 \usepackage[table,xcdraw]{xcolor}

\usepackage{float}
\usepackage{lipsum}
\usepackage{grffile}
\makeatletter
\newcommand{\distas}[1]{\mathbin{\overset{#1}{\kern\z@\sim}}}%
\newsavebox{\mybox}\newsavebox{\mysim}
\newcommand{\distras}[1]{%
  \savebox{\mybox}{\hbox{\kern3pt$\scriptstyle#1$\kern3pt}}%
  \savebox{\mysim}{\hbox{$\sim$}}%
  \mathbin{\overset{#1}{\kern\z@\resizebox{\wd\mybox}{\ht\mysim}{$\sim$}}}%
}

\usepackage{natbib}
\usepackage{authblk}

\usepackage[titles]{tocloft}

\title{\large \textbf{Topological Data Analysis (TDA) for Time Series}
\date{}}

\author[1]{Nalini Ravishanker\thanks{nalini.ravishanker@uconn.edu}}
\author[1]{Renjie Chen\thanks{renjie.chen@uconn.edu}}
\affil[1]{Department of Statistics, University of Connecticut}

\providecommand{\keywords}[1]{\textbf{\textit{Keywords}} #1}

\begin{document}
\maketitle

\begin{abstract}
The study of topology is strictly speaking, a topic in pure mathematics. However in only a
few years, Topological Data Analysis (TDA), which refers to methods of utilizing topological features in data  (such as connected components, tunnels, voids, etc.)  has gained considerable momentum. More recently, TDA is being used to understand time series.
This article provides a review of TDA for time series, with examples using R functions. Features derived from TDA are useful in classification and clustering of time series and in detecting breaks in patterns.
\end{abstract}


\section{Introduction}

Topological Data Analysis (TDA) is now an emerging area for 
analyzing complex data. TDA refers to a class of methods that garner information from topological structures in data that belong to a topological space, i.e., a mathematical space that allows for continuity, connectedness, and convergence \citep{CarlssonBulletin,harer2010}. Output from TDA may then be used for effective statistical learning about the data.
TDA combines algebraic topology and other tools from pure mathematics to allow a useful study of \textit{shape} of the data. The most widely discussed topologies of data include connected components, tunnels, voids, etc., of a topological space. Computational (or algorithmic) topology, is an overlap between the mathematical underpinnings of topology with computer science, and consists of two parts, i.e., measuring the topology of a space and persistent homology \citep{2017Chazalintroduction}. 
Using computational topology, TDA aims at analyzing topological features of data and representing these features using low dimensional representations \citep{CarlssonBulletin}. 
In particular, the space must first be represented as simplicial complexes, the Vietoris-Rips complex and the  \v{C}ech complex being the most common pathways to obtaining output to characterize the topology.

Persistent homology refers to a class of methods for measuring topological features of shapes and functions. It converts the data into simplicial complexes and describes the topological structure of a space at different spatial resolutions.   
Topologies that are more persistent are detected over a wide range of spatial scales and are deemed more likely to represent true features of the underlying space rather than sampling variations, noise, etc. Persistent homology therefore elicits persistence of essential topologies in the data and outputs the birth and death of such topologies via a persistence diagram, which is a popular summary statistic in TDA. 
Data inputs for persistent homology are usually represented as point clouds or as functions, while the outputs depend on the nature of the analysis and commonly consist of either a persistence diagram, or a persistence landscape. 
%If the data is represented as a point cloud, the 
A point cloud of data represents a sample of points from an underlying manifold and its persistent homology  approximates the topological information of the manifold. 
If data is represented as a Morse function (i.e., a smooth function on the manifold such that all critical points are non-degenerate), the persistent homology of the function is mathematically equivalent to analyzing the topological information of the manifold. % which is the core contribution in differential topology  \citep{harer2010}. 
For rigorous expositions on  algebraic topology and computational homology, see 
\cite{Munkres93a} %for algebraic topology, 
and \cite{harer2010}.  %for computational homology.

TDA has been used in cosmic web \citep{ 2013Weygaert}, shape analysis
 \citep{Carlsson2004,Chazal2009shapestable, DiFabio2011,DIFABIO20121445, Chazal2012shape,li2014chazal, Carriere2015, Bonis2016chazal}, biological data analysis \citep{DEWOSKIN2010157, Nicolau2011TopologyBD,heo2012topological, Kovacev2016,bendich2016persistent,wang2018topological}, sensor networks \citep{Silva07homologicalsensor, Silva07homologicalsensorconver, 2013Adams}, as well as other fields.

Development of TDA for time series is a relatively new and fast growing area, with many interesting applications in several different domains. 
%Applications have constructed point clouds from time series and looked at topological features. 
\cite{berwald2013automatic}  discussed the use of TDA in climate analysis. \cite{Khasawneh2015} used notions of persistence of $1$-th homology groups of point clouds (obtained via Takens's embedding) within multiple windows of time series to track the stability of dynamical systems, while \cite{Seversky2016} explored stability of various single-source and  multi-source signals. 
\cite{Perea2015SW1PerSSW} used the notion of maximum persistence of homology groups to quantify periodicity of time series. 
\cite{Pereira2015} used features derived from persistent homology to cluster populations of Tribolium flour beetles.
\cite{YuheiUmeda2017D} used topological features of one and two dimensional homology groups as inputs into %one-dimensional 
convolutional neural networks for classification of time series in three different domains, showing that their approach outperformed the baseline algorithm in each case. 
One illustration consisted of motion sensor data of daily and sports activities, 
an area also investigated using TDA by \cite{stolz2017persistent}.
\cite{truong2017exploration} as well as \cite{gidea2017topological,
gidea2018topological} and \cite{gidea2018topologicalcry} explored the use of TDA on financial time series. We discuss some of these applications in detail later in this paper.

It is well known that time series do not naturally have point cloud representations. Transformation from a time series to a point cloud is implemented through Takens's embedding 
\citep{takens1981detecting}, guaranteeing the preservation of topological properties of the time series. 
The approach consists of  transforming a time series $\{x_t, t=1,2,\ldots,T\}$,  into its phase space, i.e., a point cloud or a  set of points $\mathbf{v}_i = \{x_{i}, x_{i+\tau}, \ldots, x_{i+d \tau}\}, i=1,2,\ldots,T-d\tau$, where $\tau$ is a delay parameter and $d$ specifies the dimension of the point cloud. We discuss Taken's embedding in Section \ref{tstdapc}  and the selection of $d$ and $\tau$ in Section \ref{takens}.
TDA of time series through suitable functions is much less explored.
\cite{wang2018topological} proposed TDA on weighted Fourier series representations (Morse functions) of  electroencephalogram (EEG) data.  They used a randomness test approach to examine properties of the proposed method and show its robustness to different transformations of the data.  
TDA of time series through sublevel set filtration of functions 
is discussed in Section \ref{FunctionTS}.

The format of this paper follows. Section \ref{PointCloudTS} provides a review of TDA from point clouds and then describes TDA for time series via the Takens's embedding method. Section \ref{FunctionTS} provides a review of persistent homology on functions and then describes TDA for time series analysis starting from  second-order spectra or Walsh Fourier transforms. Section \ref{FeaturesTDA} discusses constructing TDA based features which are then used in learning about time series, with  applications on classification, clustering and detecting changes in patterns. Section \ref{summary} gives a discussion and summary.

\section{Persistent Homology Based on Point Clouds} \label{PointCloudTS}

In the section, we first describe persistent homology of a manifold starting from point cloud data, followed by its construction and use in time series analysis using Takens's embedding.

\subsection{Point Clouds to Persistence Diagrams - A Basic Review}  \label{tda_pc}

Starting from a point cloud, we show the procedure to elicit topological features of data.
Denote the point cloud as $\mathcal{P} = \{\mathbf{v}_i: i=1,2,\ldots, N\}$, where $\mathbf{v}_i \in \mathcal{R}^d$. When $d=2$, the points lie on the plane.
Let  
%$D_{E}(\mathbf{v}_{i},\mathbf{v}_{j})$ denote the Euclidean distance between $\mathbf{v}_i$  and $\mathbf{v}_j$, for $i,j=1,\ldots,N$, and let 
$\mathbf{D}_E =\{D_{E}(\mathbf{v}_{i},\mathbf{v}_{j} )\}$ be the  $N\times N$ matrix of Euclidean distances, for $i,j=1,\ldots,N$.
For each $\mathbf{v}_{i} \in \mathcal{P}$, let  $\mathbf{B}_\lambda (\mathbf{v}_i) = \{\mathbf{x}:  D_{E}(\mathbf{x}, \mathbf{v}_i)\leq \lambda/2, \mathbf{x} \in \mathcal{R}^d\}$ denote a closed ball with radius $\lambda/2$; here, 
$0 \le \lambda \le U$, where  the upper-bound $U$ is usually pre-determined as the maximum of the distances in 
$\mathbf{D}_E$.
%where $\lambda$ is also the parameter using for the filtration. 
A Vietoris-Rips simplex \citep{harer2010} corresponding to a given $\lambda$ is defined as the 
set of points $\mathcal{P}_{V}(\lambda) \subset \mathcal{P}$ 
such that any points $\mathbf{v}_{i_1}, \mathbf{v}_{i_2}$ in $\mathcal{P}_{V}(\lambda)$ satisfy 
$D_{E}(\mathbf{v}_{i_1}, \mathbf{v}_{i_2}) \leq \lambda$, $1\leq i_1, i_2 \leq N$.
For a given $\lambda$ value, a simplicial complex  $\tilde{\kappa} (\lambda)$  denotes the set of Vietoris-Rips simplexes such that for any two Vietoris-Rips simplexes $\mathcal{P}_{V}^{(1)}(\lambda), \mathcal{P}_{V}^{(2)}(\lambda) \in \tilde{\kappa} (\lambda)$, %\rightarrow 
we have (i) $\mathcal{P}_{V}^{(1)}(\lambda) \cap \mathcal{P}_{V}^{(2)}(\lambda) \in \tilde{\kappa} (\lambda)$ and (ii) 
%for any subset of a simplex, 
if $\mathcal{P}' \subset \mathcal{P}_V^{(1)}$,
%\rightarrow 
then $\mathcal{P}' \in \tilde{\kappa} (\lambda)$. 

A simplicial complex consisting of $(\tilde{p} + 1)$ points (from different Vietoris-Rips simplexes) is a $\tilde{p}$-dimensional simplicial complex. 
In algebraic topology, $\tilde{p}$ is at most $N-1$ when the point cloud had $N$ points.  
The topology of the point cloud  is studied through the topology of the simplicial complexes, denoted by    
$\{\tilde{\alpha}_{\tilde{p},k}: k=1,2,\ldots,k_{\tilde{p}}\}$, and 
$\tilde{\alpha}_{\tilde{p},k}$ is a homology group, consisting of a set of $\tilde{p}$-dimensional simplicial complexes which   
are homomorphic. 
%Here, $k_{\tilde{p}}$ is the number of homology groups among all $\tilde{p}$-dimensional simplicial complexes. 
For the theory and computation of homomorphisms, refer to 
\cite{Munkres93a, carlsson2014topological} and 
\cite{harer2010}. 
As the parameter $\lambda$ gradually increases,
the birth and death of homology groups $\{\tilde{\tau}_{\tilde{p},k}=(\lambda_{\tilde{p},k,1}, \lambda_{\tilde{p},k,2}): k=1,2,\ldots,k_{\tilde{p}}\}$ are recorded in the persistence diagram. 
A $\tilde{p}$-th Betti number of $\lambda$ is the number of $\tilde{p}$-th homology groups at $\lambda$, denoted as $k_{\tilde{p}}^{(\lambda)}$. 

% or{red}{RC: phrase below does not make sense? Are you denoting the $\tilde{p}$-th Betti number of $\lambda$ as $k_{\tilde{p}}^{(\lambda)}$? Why not just say this?} namely counting existing $\tilde{\alpha}_{\tilde{p},k}$ at $\lambda$ as $k_{\tilde{p}}^{(\lambda)}$.

Computation of the topological features are summarized in the following steps.

\begin{description}
\item [Step 1.] Compute the Euclidean distance matrix $\mathbf{D}_E = \{D_{E}(\mathbf{v}_{i_1}, \mathbf{v}_{i_2})\}$ for $i_1, i_2\in\{1,\ldots, N\}$; this is the default distance for a point cloud in \texttt{R-TDA}.

\item [Step 2.] Construct birth and death of homology groups for increasing values of $\lambda$. For each $\lambda$, compute $\tilde{\alpha}_{\tilde{p},k}$ from $\tilde{\kappa} (\lambda)$ using closed balls $\mathbf{B}_\lambda (\mathbf{v}_i)$ of $\mathbf{v}_i$ with radius $\lambda/2$. 
If an elder topology $\tilde{\alpha}_{\tilde{p},k_1}$ and a younger one $\tilde{\alpha}_{\tilde{p},k_2}$ merge into a single $\tilde{\alpha}_{\tilde{p},k}$ at some $\lambda$, $\tilde{\alpha}_{\tilde{p},k_1}$ would become $\tilde{\alpha}_{\tilde{p},k}$ and $\tilde{\alpha}_{\tilde{p},k_2}$ would die.
%at  $\lambda$.  

\item [Step 3.] The persistence diagram is an output of the set of points representing birth-death of homology groups from the point cloud and is denoted as  $\tilde{\Omega} = \{\tilde{\tau}_{\tilde{p},k}=(\lambda_{\tilde{p},k,1}, \lambda_{\tilde{p},k,2}): \tilde{p}=0,1, \ldots;k=1,2,\ldots,k_{\tilde{p}}\}$.  We plot  $\lambda_{\tilde{p},k,1}$ on the $x$-axis and $\lambda_{\tilde{p},k,2}$ on the $y$-axis \citep{harer2010}. 
\end{description}
\vskip1em

\noindent \textbf{Example 2.1. Point Cloud to Persistence Diagram.} We illustrate construction of the persistence diagram for a point cloud with $N=60$ points sampled from  the  unit circle  $x_1^2+x_2^2=1$:  
\begin{verbatim}
set.seed(1); PC <- circleUnif(n = 60, r = 1)
plot(PC, main = "(a)")
\end{verbatim}
The point cloud is shown in Figure \ref{pointcloudpersist}(a).  We expect to see a total of $k_0=60$ values of $\tilde{\alpha}_{0,k}$ and $k_1 =1$ value of $\tilde{\alpha}_{1,k}$.
We use the function \texttt{ripsDiag}  from the  \texttt{R-TDA} package for constructing  the persistence diagram
\citep{Fasy2014IntroductionTT}.
In the R code chunk shown below, \texttt{PC} denotes the input point cloud, \texttt{maxdimension} is the maximum dimension $\tilde{p}$ of points $\tilde{\tau}_{\tilde{p},k}$ to be calculated, and \texttt{maxscale} is the maximum value that the filtration parameter $\lambda$ can assume. We set \texttt{maxdimension} to be 1. %, especially for large point clouds (i.e.,  $N>100$).
The default \texttt{dist} is the Euclidean distance. The output \texttt{pers.diag.1} returns the persistence diagram, as a matrix with three columns which summarize topological features of the point cloud. 
\begin{verbatim}
(pers.diag.1 <- ripsDiag(X=PC, maxdimension = 1, maxscale = max(dist(PC))) )
$diagram
      dimension     Birth       Death
 [1,]         0 0.0000000 1.999999902
 [2,]         0 0.0000000 0.306978455
 [3,]         0 0.0000000 0.245260715
 .....
 [60,]         0 0.0000000 0.027923092
 [61,]         1 0.3190835 1.737696840
\end{verbatim}
The first row  with $\tilde{\tau}_{0,1} = (0, 2.00)$ 
in the output records that there is a $0$-th homology group (connected component) whose birth happens at $\lambda=0$ and whose death happens at about $\lambda=2.00$. 
The second row with $\tilde{\tau}_{0,1} = (0, 0.31)$ records that the second connected component is born at $\lambda=0$ and is dead at $\lambda=0.31$, etc. We see that all $0$-th homology groups have birth time $0$,
and all $N=60$ points start as connected components. 
These 60 connected components are shown in decreasing order of persistence (slower death). 
Row 61 with $\tilde{\tau}_{1,1} = (0.32, 1.74)$ describes the birth and death of a $1$-th homology group (tunnel) at $\lambda=0.32$ and  $\lambda=1.74$ respectively.  
Figure \ref{pointcloudpersist}(b) corresponds to the filtration parameter $\lambda = 0$ and is obtained using this code:
\begin{verbatim}
plot(x=1,y=1,type="n",ylim=c(0,2),xlim=c(0,2),ylab="death",xlab="birth",main="(b)")
abline(v = 0, lty = 2)
\end{verbatim}
The dashed vertical line 
indicates the birth time of connected components; the plot has no points because none of the connected components has died.
Figure \ref{pointcloudpersist}(c) corresponds to the point cloud when $\lambda = 0.1$:
\begin{verbatim}
plot(PC, pch = 16, cex = 5, col = "blue", main = "(c)")
\end{verbatim}
The blue
balls $\mathbf{B}_\lambda (\mathbf{v}_i)$ around each point enlarge and connect with others, resulting in fewer connected components. 
The black 
dots in Figure \ref{pointcloudpersist}(d) denote the birth-death times of the merged connected components which have died before $\lambda=0.1$:
\begin{verbatim}
death.time = sort(pers.diag.1$diagram[pers.diag.1$diagram[, 1]==0, 3])
plot(x = rep(0, sum(death.time<=0.1)), y = death.time[which(death.time<=0.1)], 
    ylim = c(0, 2), xlim = c(0, 2), ylab = "death", xlab = "birth", main = "(d)")
abline(v = 0, lty = 2); abline(h = 0.1, lty = 2)
\end{verbatim}
When $\lambda = 0.32$ in Figure \ref{pointcloudpersist}(e), all points connect together and a tunnel emerges, which is the white area surrounded by the blue 
circle:
\begin{verbatim}
plot(PC, pch = 16, cex = 12, col = "blue", main = "(e)")
\end{verbatim}

\begin{figure}[H]
\centering
\includegraphics[width=\textwidth]{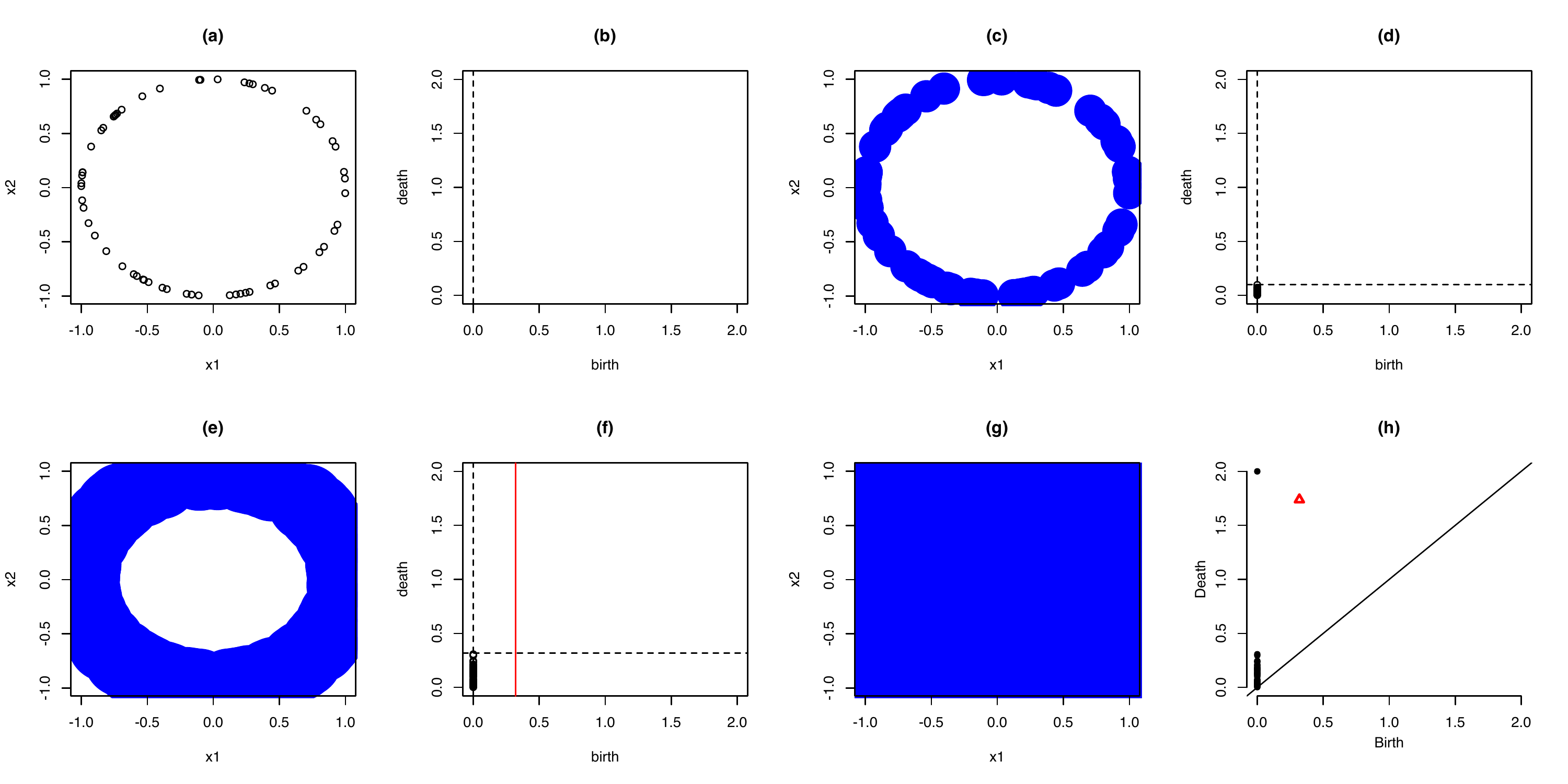}
\caption{Persistence diagram corresponding to a point cloud. (a) shows the raw point cloud and (h) shows the persistence diagram. (c), (e) and (g) are intermediate steps for the filtration by varying $\lambda$, while (b), (d) and (f) are intermediate steps for constructing the persistence diagram.}
\label{pointcloudpersist}
\end{figure}

The birth time of this tunnel is recorded as $\lambda = 0.32$, which is shown as the red dashed line 
in Figure \ref{pointcloudpersist}(f). Further, there are more black 
dots in this figure since there are more connected components that have died before $\lambda=0.32$:
\begin{verbatim}
plot(x = rep(0, sum(death.time<=0.32)), y = death.time[which(death.time<=0.32)], 
ylim = c(0, 2), xlim = c(0, 2), ylab = "death", xlab = "birth", main = "(f)")
abline(h = pers.diag.1$diagram[pers.diag.1$diagram[, 1]==1, 2], lty = 2)
abline(v = 0, lty = 2); abline(v = 0.32, col = "red", lty = 1)
\end{verbatim}
When $\lambda$ reaches its maximum value of $2$ (which is the largest value  %$\max_{i_1,i_2} 
in $D_{E}(\mathbf{v}_{i_1}, \mathbf{v}_{i_2})$), the algorithm stops and outputs the persistence diagram (see Figure \ref{pointcloudpersist}(h))
which finally shows the birth-death times for all connected components (the dots) 
and the tunnel (the red  triangle):
\begin{verbatim}
plot(PC, pch = 16, cex = 40, col = "blue", main = "(g)")
plot(pers.diag.1$diagram, main = "(h)")
\end{verbatim}
\vskip1em

\texttt{R-TDA} also supports construction of a persistence diagram given an arbitrary distance matrix as input, as shown in the example below.
\vskip1em

\noindent \textbf{Example 2.2. Distance Matrix to Persistence Diagram.} 
The input  to \texttt{ripsDiag} can be 
a distance matrix computed from the point cloud generated in Example 2.1. Here, we use the default \texttt{dist="euclidean"}. Other options are 
\texttt{"manhattan"}, \texttt{"maximum"}, etc.

\begin{verbatim}
dist.PC <- dist(PC)
(pers.diag.2=ripsDiag(X=dist.PC,dist="arbitrary",maxdimension=1,maxscale=max(dist.PC)))
$diagram
      dimension     Birth       Death
 [1,]         0 0.0000000 1.999999902
 [2,]         0 0.0000000 0.306978455
 [3,]         0 0.0000000 0.245260715
 .....
 [60,]         0 0.0000000 0.027923092
 [61,]         1 0.3190835 1.737696840
\end{verbatim}

\subsection{Distances Between Persistence Diagrams}  \label{TDAdist}

Two distance metrics are commonly used  to quantify the dissimilarity between two persistence diagrams $\tilde{\Omega}_{1}$ and  $\tilde{\Omega}_{2}$ , the Wasserstein distance and the bottleneck distance
\citep{Mileyko2011}. We define these distances and describe their computation using the R-TDA package.

The $q$-Wasserstein distance  between two persistence diagrams is defined by
\begin{equation}  \label{wasserstein}
\mathbf{W}_{q, \tilde{p}} (\tilde{\Omega}_1, \tilde{\Omega}_2) = \big[\inf_{\eta:\tilde{\Omega}_1\rightarrow \tilde{\Omega}_2} \sum_{\tilde{\tau}_{\tilde{p},k}\in \tilde{\Omega}_1} |\tilde{\tau}_{\tilde{p},k}-\eta(\tilde{\tau}_{\tilde{p},k})|_\infty^q \big]^{1/q}, q = 1,2,\ldots,
\end{equation}
where $\tilde{p}$ is referred to as its dimension and $q$ is its degree.
When $q=\infty$, (\ref{wasserstein}) is the bottleneck distance of dimension $\tilde{p}$ defined by
\begin{equation} \label{bottleneck}
\mathbf{W}_{\infty, \tilde{p}} (\tilde{\Omega}_1, \tilde{\Omega}_2) = \inf_{\eta:\tilde{\Omega}_1\rightarrow \tilde{\Omega}_2} \sup_{\tilde{\tau}_{\tilde{p},k}\in \tilde{\Omega}_1} |\tilde{\tau}_{\tilde{p},k}-\eta(\tilde{\tau}_{\tilde{p},k})|_\infty.
\end{equation}
The bottleneck distance is obtained by minimizing the largest distance of any two corresponding points of diagrams, over all bijections between $\tilde{\Omega}_{1}$ and $\tilde{\Omega}_{2}$  
and is less sensitive to details in the diagrams. 
\vskip1em

\noindent \textbf{Example 2.3. Wasserstein and Bottleneck Distances.}  
Let $\tilde{\Omega}_1$ be \texttt{pers.diag.2}, the persistence diagram obtained in Example 2.2. Let $\tilde{\Omega}_2$ be \texttt{pers.diag.3}, the persistent diagram we construct from a different point cloud from the same unit circle. 
We use \texttt{R-TDA} to compute the Wasserstein distance with $q=1$ (denoted by the argument \texttt{p=1} below):
\begin{verbatim}
set.seed(2); PC2 <- circleUnif(n = 60, r = 1)
pers.diag.3 <- ripsDiag(X=PC2, maxdimension = 1,maxscale = max(dist(X))$diagram
wasserstein(pers.diag.2, pers.diag.3, p=1, dimension = 0)
1.034579
\end{verbatim}

The function \texttt{bottleneck} enables us to compute the bottleneck distance between the two persistence diagrams. 
The Wasserstein distance is larger than the bottleneck distance since  the former measures more detailed difference between the diagrams. 
\begin{verbatim}
bottleneck(pers.diag.2, pers.diag.3, dimension = 0)
0.06954618
\end{verbatim}

It is important to construct persistence diagrams using the same distance functions 
\citep{JMLRChazal2018}, as we show below.
For instance, we can construct a persistence diagram \texttt{pers.diag.4} for the point cloud in Example 2.1 using the  Manhattan distance 
($D_M (\mathbf{v}_i, \mathbf{v}_j)=\sum_{\ell=1}^{d} |v_{i,\ell}-v_{j,\ell}|, 1\leq i,j\leq N$) instead of the Euclidean distance. 
\begin{verbatim}
dist.PC.man <- dist(PC, method = "manhattan"); max.dist=max(dist.PC.man)
pers.diag.4=ripsDiag(X=dist.PC.man,dist="arbitrary",maxdimension=1,
maxscale=max.dist)$diagram
wasserstein(pers.diag.2, pers.diag.4, p=1, dimension = 0)
2.145083
bottleneck(pers.diag.2, pers.diag.4, dimension = 0)
0.8279462
\end{verbatim}

\subsection{TDA of Time Series via Point Clouds} \label{tstdapc}

Time series do not naturally have point cloud representations, and are transformed to point clouds using 
Takens's Embedding Theorem \citep{takens1981detecting} 
before we can do TDA as discussed in Section  \ref{tda_pc}.  
This approach has been used in the literature mostly for quantifying periodicity in time series \citep{Perea2015SW1PerSSW}, clustering time series \citep{Seversky2016}, classifying time series \citep{YuheiUmeda2017D}, or finding early signals for critical transitions \citep{gidea2017topological,gidea2018topological}.  Takens’s embedding guarantees the preservation of topological properties of a time series but not  its geometrical properties. 

\subsubsection{Takens's Delay Embedding for Time Series} \label{takens}

Let $\{x_t,  t=1,2,\ldots,T\}$ denote an observed  time series. We use Takens's embedding to convert the time series into a point cloud with points 
$\mathbf{v}_i = (x_i, x_{i+\tau}, \ldots, x_{i+(d-1)\tau})^{\prime}$, where $d$ specifies the dimension of the points and $\tau$ denotes a delay parameter.
For example, if $d=2$ and $\tau=1$, then, $\mathbf{v}_i = (x_i, x_{i+1})^{\prime}$, whereas if
$d=15$ and $\tau=2$, $\mathbf{v}_i = (x_i, x_{i+2},\ldots, x_{i+28})^{\prime}$. 
Both $d$ and $\tau$ are unknown and must be determined in practice.
\vskip1em

\noindent \textbf{Choice of $\tau$.}
Researchers have used different approaches for choosing  
the delay parameter $\tau$.  It may be selected as the smallest time lag $h$ where the sample 
autocorrelation function (ACF) $\hat{\rho}_h$ becomes insignificant, i.e., smaller in absolute value than the critical bound  $\frac{2}{\sqrt{T}}$ \citep{Khasawneh2015}. 
\cite{truong2017exploration} also used the ACF, but in a slightly different way. He chose $\tau$ as the 
smallest lag for which  $(\hat{\rho}_{\tau}-\hat{\rho}_{\tau-1})/\hat{\rho}_{\tau}>1/e$ and $\hat{\rho}_{\tau}<\frac{2}{\sqrt{T}}$. 
\cite{Pereira2015} determined $\tau$ using the first minimum of the auto mutual information (the mutual information between the signal and its time delayed version).

\vskip1em

\noindent \textbf{Choice of $d$.}
\cite{truong2017exploration} and \cite{Khasawneh2015} used the false nearest neighbor method 
\citep {kennel1992determining} to determine the embedding dimension $d$ as the integer such that 
the nearest neighbors of each point 
in dimension $d$ remain nearest neighbors in dimension $d+1$, and the distances between them also remain about the same. Alternately, an \texttt{R} function \texttt{false.nearest} in the package \texttt{tseriesChaos} which implements an approach due to \cite{hegger1999practical} may be used. Some authors \citep{Pereira2015,Seversky2016}, simply assume $d$ to be $2$ or $3$, while Perea has suggested the use of  
$d=15$ on time series after a cubic spline interpolation (see Section \ref{SW1PerS}).

\vskip1em

The choice of $d$ and $\tau$ then determine the number of points $N$ in the point cloud. In Example 2.4, we illustrate one approach for constructing a point cloud from pure periodic signals with no noise and then obtaining a persistence diagram. In Example 2.5, we discuss another approach described in \cite{Perea2015} for noisy time series, when the focus is on finding series with the same periodicity.
\vskip1em

\noindent \textbf{Example 2.4. Pure Signals to Persistence Diagrams.}
We generate point clouds from three periodic cosine signals of length $T=480$ with periods 12, 48, and 96 respectively, and then construct their persistent diagrams. We set $d=2$, and use the ACF method discussed above to choose $\tau$.
We show R code for the time series \texttt{ts1}:
\begin{verbatim}
per1=12;ts1 = cos(1:T*2*pi/per1);d=2; 
tau <- which(abs(acf(ts.ex, plot = F)$acf) < 2/sqrt(T))[1]-1
PC=t(purrr::map_dfc(1:(T-(d-1)*tau+1),~ts.ex[seq(from=.x, by=tau, length.out=d)]))
diag=ripsDiag(PC, maxdimension=1, maxscale=max(dist(PC)))
ts.plot(ts.ex);plot(PC,xlab ="x1",ylab="x2",main="PC");plot(diag$diagram) 
\end{verbatim}

\begin{figure}[H]
\centering
\includegraphics[width=0.9\textwidth]{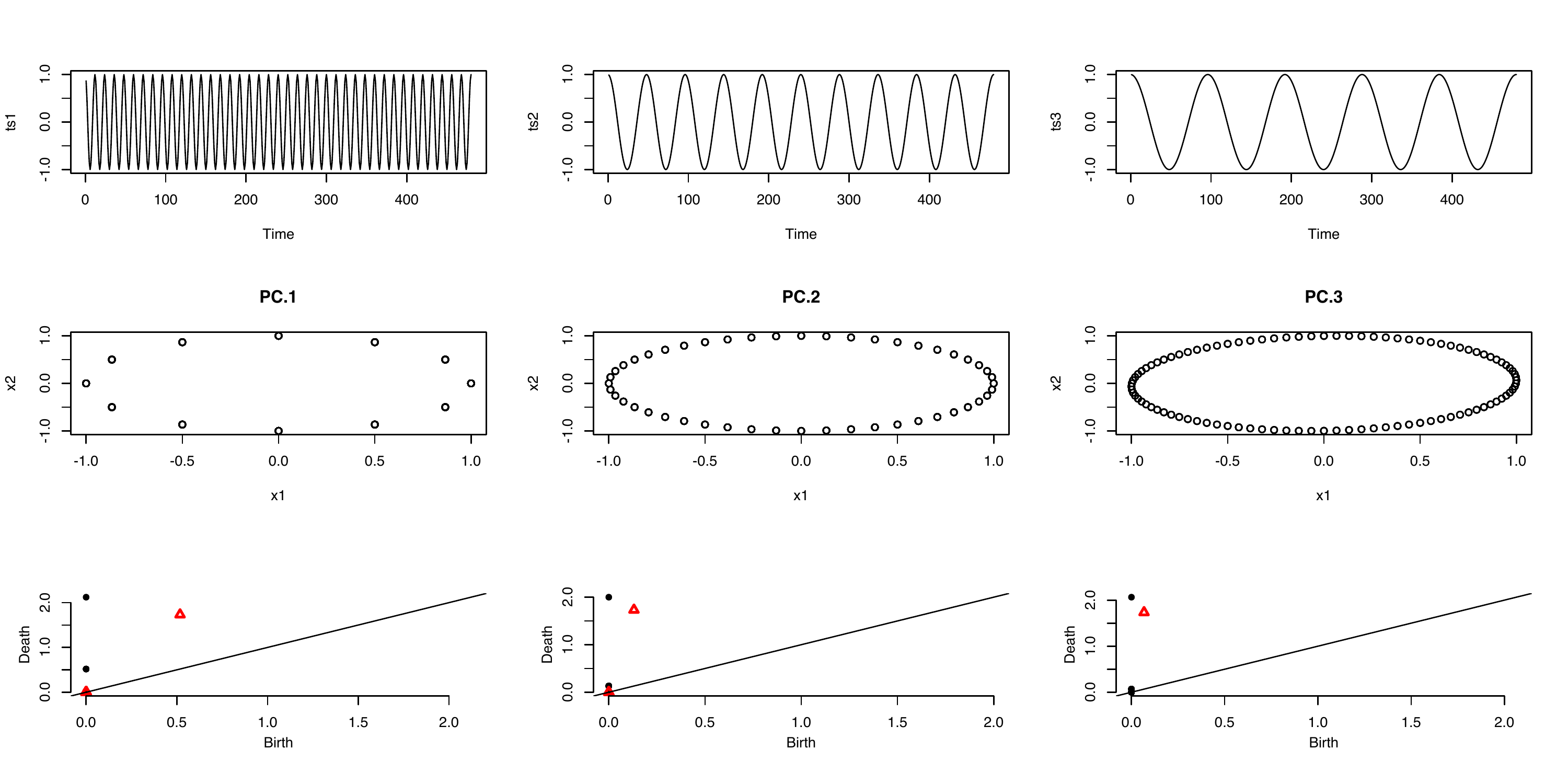}
\caption{Pure Periodic Signals to Persistence Diagrams.}
\label{tsexSimple24}
\end{figure}

In Figure \ref{tsexSimple24}, the top row shows the signals, the middle row shows the point clouds and the bottom row shows the persistence diagrams.
The black dots represent the birth-death of $0$-th homology groups and their persistence shows the dispersion of the points in the point cloud. When there are more black dots close to the diagonal, the point cloud is more dispersed. Particularly, the point cloud PC.3 from the time series with period 96 have points close to each other compared with PC.1, so that it has more black dots in the persistence diagram closer to the diagonal. 

The red triangles represent the birth-death of $1$-th homology groups, indicating circles in the point cloud.  
The red triangle from the time series with period 96 is further away from the diagonal compared to the series with  period 12, and thus has longer  persistence in the $1$-th homology group. Seeing a circle indicates that the time series is periodic.
This is in contrast to the persistence diagram for the same time series based on sublevel set flitration of a function, as discussed in Example 3.2.

Pairwise bottleneck distances between the three persistence diagrams are shown in Table \ref{BnDistPDTF}, computed using code as shown below:
\begin{verbatim}
round(bottleneck(diag1$diagram, diag2$diagram, dimension = 0), digits = 2)
\end{verbatim}
To study the effect of $d$, we repeat the computations for $d=3$ and $d=15$ and also show all pairwise bottleneck distances in the table. While the values of the distances between the diagrams change as $d$ changes, the relative behavior is 
preserved, independent of $d$.  
Specifically, the bottleneck distances between
$\tilde{\Omega}_1$ and
$\tilde{\Omega}_2$ and between $\tilde{\Omega}_1$ and
$\tilde{\Omega}_3$ are larger than the distance  between $\tilde{\Omega}_2$ and
$\tilde{\Omega}_3$ in the  0-th and 1-th homology groups.

\begin{table}[H]\centering
\caption{Pairwise Bottleneck Distances Between Persistence Diagrams for Different $d$.}\label{BnDistPDTF}
\begin{tabular}{|c|rrr|rrr|rrr|}
\hline
& \multicolumn{3}{c|}{$d=2$} & \multicolumn{3}{c|}{$d=3$}& \multicolumn{3}{c|}{$d=15$}\\\hline
& (1,2) & (1,3) & (2,3)& 
(1,2) & (1,3) & (2,3) & 
(1,2) & (1,3) & (2,3)     \\\hline
0th & 0.26 & 0.26 & 0.07&
0.36 & 0.36 & 0.09 &
0.73 & 0.73 & 0.18
\\\hline
1th & 0.39 & 0.45 & 0.06 &
0.53 & 0.63 & 0.09 &
1.09 & 1.28 & 0.19
\\\hline
\end{tabular}
\end{table}

\subsubsection{Point Cloud Construction using SW1PerS}\label{SW1PerS}

The SW1PerS (Sliding Windows and 1-Persistence Scoring) method is an alternate, more comprehensive approach 
%for point cloud construction 
proposed by \cite{Perea2015} to detect periodicity from noisy time courses whose underlying signals may have different shapes. 
The approach addresses the following items. 
\vskip1em

\noindent \textbf{Denoising.}
The approach considers two types of denoising that are left as options to the user. The first type smooths the raw time series by a moving average in order to make it easier to detect the signal.  
The second type is a moving average on the point cloud. 
As an alternative to moving averaging,   
\cite{Pereira2015} used the Empirical Mode Decomposition (EMD) \citep{huang1998empirical} on the raw time series.
\vskip1em

\noindent \textbf{Spline Interpolation.}
The spline interpolation allows handling  unevenly spaced time series, or time series with low temporal resolution.
\vskip1em

\noindent \textbf{Point Cloud Standardization.}
Standardization helps with 
signal dampening and to make the procedure amplitude blind.
%so that points lie in a planar circle of radius 1.
\vskip1em

\noindent The pipeline for this approach is described in the following steps.

\begin{description}  \label{ntsprocedure}

\item [Step 0.] Optionally  \citep{Perea2015SW1PerSSW}, denoise the observed  time series $\{x_t, t=1,2,\ldots,T\}$
using a simple moving average 
whose window size is no higher than one third of the selected dimension $d$.   
They recommended an embedding dimension of $d=15$ and $N=201$ as the size of the point cloud; then  $T_1 = N+d= 216$.  

\item [Step 1.] For selected values of $d$ and $\tau$ (see below), create a point cloud from the (possibly denoised) time series using Steps 1.1 and 1.2.
\begin{itemize}

\item [Step 1.1.] Recover a continuous function $g: [0, 2\pi] \rightarrow \mathcal{R}$ by 
fitting a cubic spline to the denoised time series $\{x_t, t=1,2,\ldots,T\}$.   

\item [Step 1.2.] Using values $g(t_1), g(t_2), \ldots, g(t_{T_1})$ from the continuous spline fit $g(.)$ at evenly spaced time points $0= t_1\leq t_2\leq \ldots\leq t_{T_1}=(T_1 -1)\tau=2\pi$, construct 
a point cloud with 
$N = T_1 -d$ points $\mathbf{v}^{(0)}_t=(g(t), g(t+\tau), \ldots, g(t+(d-1)\tau))' \in \mathcal{R}^d, t=0, \tau,\ldots,2\pi-(d-1)\tau$ and so $\tau=\frac{2\pi}{N+d-1}$. 
\end{itemize}
\item [Step 2.] Pointwise Point cloud standardization:
\begin{gather}
\mathbf{v}_t = \frac{\mathbf{v}^{(0)}_t-\bar{v}^{(0)}_t \mathbf{1}}{||\mathbf{v}^{(0)}_t-\bar{v}^{(0)}_t \mathbf{1}||};\quad \bar{v}^{(0)}_t = \sum_{i=1}^{d}v^{(0)}_{t,i}/d,\quad
||\mathbf{v}^{(0)}_t-\bar{v}^{(0)}_t \mathbf{1}|| = \sqrt{\sum_{i=1}^d (v^{(0)}_{t,i}-\bar{v}^{(0)}_{t})^2},
\end{gather}
where $\mathbf{v}^{(0)}_t=(v^{(0)}_{t,1},v^{(0)}_{t,2},\ldots, v^{(0)}_{t,d})'$ and  $\mathbf{1}$ is the $d$-dimensional vector of 1's.

\item [Step 3.] Construct the persistence diagram from the point cloud as shown in Section \ref{tda_pc}. 
\end{description}

\noindent

This method is powerful for detecting periodicity in time series.
To develop a score for quantifying the periodicity,  \cite{Perea2015} first found the longest persistence of 
the birth-death of the $1$-th homology groups  $(\lambda_{1,k_M,1}, \lambda_{1,k_M,2})$, where $k_M =\mbox{arg}\max_{k}(\lambda_{1,k,2}-\lambda_{1,k,1})$ is chosen to indicate maximum persistence, and used it to compute
$$\mathcal{S} = 1- \frac{\lambda_{1,k_M,2}^2-\lambda_{1,k_M,1}^2}{3}.$$  
Since $0 \leq \lambda_{1,k_M,1} \leq \lambda_{1,k_M,2} \leq \sqrt{3}$, for periodic (nonperiodic) time series, the score is close to zero (one).
%%%%%%%%%%%%%%%%%%%%%%%%%%%%%%%%%%%%%%%%%
%
The \texttt{R} code for implementing Step 1-Step 3 for Case 1 is shown below (the code for other cases is similar):

\begin{verbatim}
x.ts = ts1;  d=15; N=201; T1 = 216;
x.ts <- pracma::movavg(x.ts, 5, type = "s") #step 0
sp.ts <- stats::spline(1:T*2*pi/T, x.ts, n=T1)$y #step 1.1
PC <- plyr::ldply(map(1:N, ~sp.ts[.x:(.x+d-1)]))#step 1.2
X.PC=t(apply(PC,1,FUN=function(x){(x-mean(x))/sqrt(sum((x-mean(x))^2))})) #step2
diag <- ripsDiag(X=x.PC, maxdimension = 1, maxscale = sqrt(3))#step 3
\end{verbatim}

The main contributions of \cite{Perea2015} are the use of extensive simulation studies to show
that topological features of time series are largely the same under various non-sinusoidal shapes 
as well as under differences in amplitude, phase, mean, frequency, or trend.  
The results may be affected by a differences in noise variances, as well as by the shapes of the noise and signal distributions.

\vskip1em

\begin{figure}[H]
\centering
\includegraphics[height = 0.7\textheight,width=0.85\textwidth]{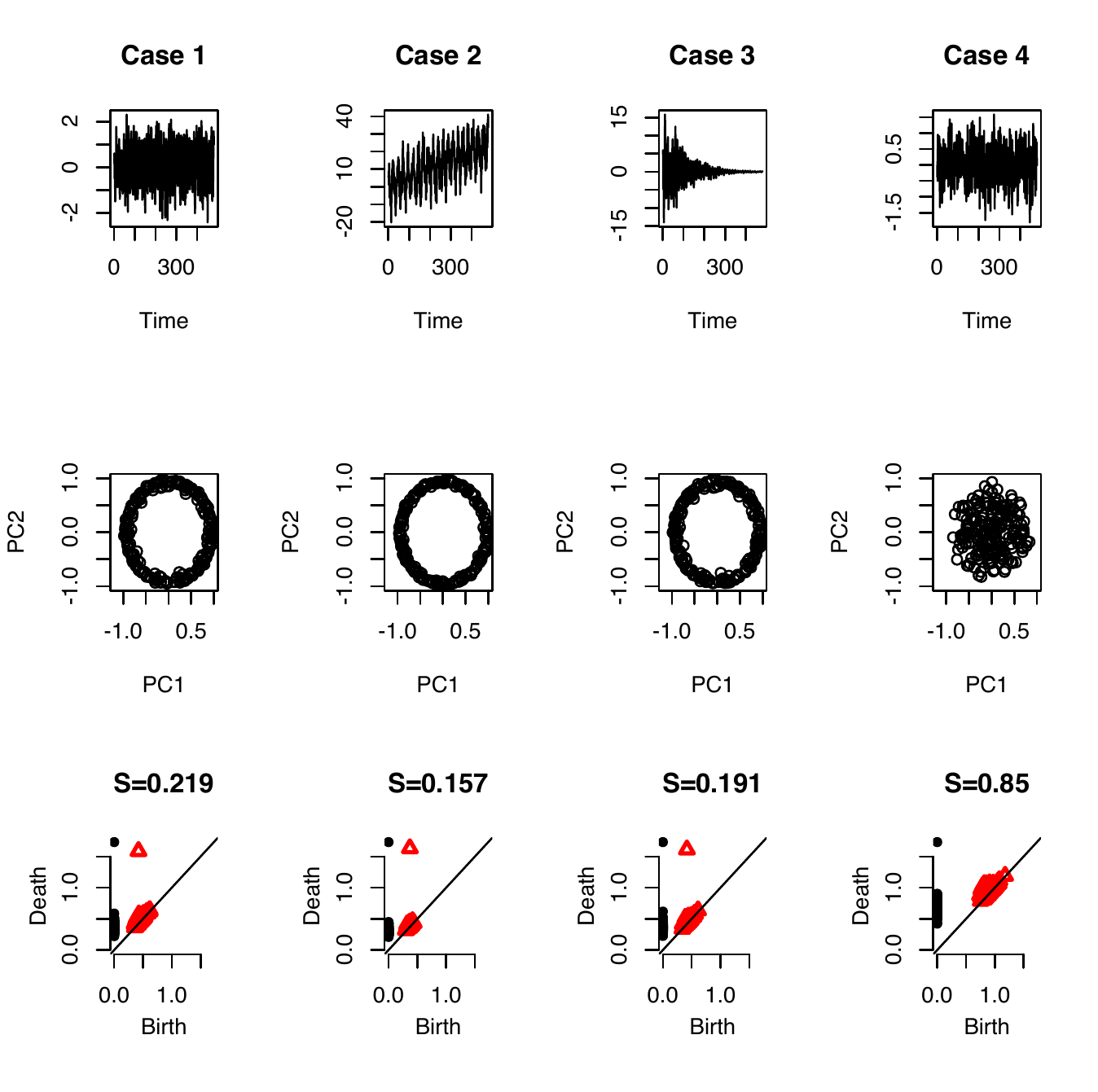}
\caption{Persistence Diagrams of Periodic Time Series with Different Shapes}
%\caption{Periodic time series with different shapes, constructed point clouds and persistence diagrams.}
\label{UseSW1PerStrendNoisePDsept12}
\end{figure}

\noindent \textbf{Example 2.5. Using SW1PerS for Periodicity Quantification.} 
We generate white noise 
$\epsilon_t$ with variance $\sigma^2_\epsilon=0.64$, and then generate
periodic time series signals each of length $T$ ($=480$), denoted by \texttt{ts1, ts2, ts3, ts4}, and shown in the top row of Figure \ref{UseSW1PerStrendNoisePDsept12}. 

\begin{itemize}
\item Case 1: $x_t=\cos(2\pi t/12)+\epsilon_t$, %\epsilon_t\sim N(0, 0.64)$
\item Case 2: $x_t=0.05t+10(\cos(2\pi (t-\varphi)/48)+\epsilon_t)$, 
%\epsilon_t\sim N(0, 0.64)$
\item Case 3: $x_t=10(\cos(2\pi t/12)+\epsilon_t)\exp{(-0.01t)}$, 
%\epsilon_t\sim N(0, 0.64)$
\item Case 4: $x_t  =\epsilon_t$
%\sim N(0, 0.64)$
\end{itemize}

The series \texttt{ts2} differs from \texttt{ts1} in frequency, phase, and linear trend, whereas  \texttt{ts3} only differs from \texttt{ts1} in shape and amplitude; \texttt{ts4} is the white noise series. 
Figure \ref{UseSW1PerStrendNoisePDsept12} shows that the method in \cite{Perea2015} is insensitive to different periodicities. The top row shows the simulated time series under the four cases. The middle row presents a two dimensional view  of the point clouds constructed  with  $d=15$ and $\tau=1$, via the first two principal components.
The bottom row shows the persistence diagrams, along with the periodicity score $S$.
For periodic (nonperiodic) time series, the score is close to zero (one), so that the scores for Case 1 to Case 3 are close to 0, while Case 4 has a score close to 1. 
 
\vskip1em

\section{Persistent Homology Based on Functions}\label{FunctionTS}

We first give a basic review of TDA based on functions, followed by the use of frequency domain representations of time series as starting points for TDA.

\subsection{Function to Persistence Diagram - A Basic Review}\label{tda_fun}

When data is in the form of  a continuous function 
$f:\mathcal{R}^d \rightarrow \mathcal{R}$,  
or can be converted to such a function, TDA using sublevel set filtration is carried out 
by discretizing the function into grids and then implementing computational homology on the discretized function. 
Suppose the components of the function are  $\mathbf{z}=(\ell_1\delta,\ell_2\delta,\ldots,\ell_d \delta)$, for $\ell_1, \ell_2, \ldots, \ell_d=0,\pm 1,\pm 2,\ldots$, where $d >0$ and $\delta >0$.  The sublevel set of the function is
defined as 
\begin{equation}  \label{sublevelset}
L_\lambda (f) = \{\mathbf{z}: f(\mathbf{z})\leq \lambda, \mathbf{z}\in \mathcal{R}^d\},
\end{equation}
where 
$0\leq \lambda\leq \max_{\mathbf{z}} f(\mathbf{z})$. 
Define a simplex as a set of components  in $L_\lambda (f)$  which are ``neighbors'', i.e., $\mathbf{z}_1, \mathbf{z}_2\in L_\lambda (f)$, and 
$|z_{1, j}-z_{2,j}| \leq\delta, j=1,2,\ldots, d$.

Recall from Section \ref{tda_pc} that a simplex with $(\tilde{p}+1)$ components is called a $\tilde{p}$-simplex. 
Since only adjacent points on the grid can be neighbors, the function $f(\mathbf{z})$ can admit at most a $d$-simplex, so that  
$\tilde{p}\leq d-1$ \citep{harer2010}.
When $d=1$, there are only connected components, 
whose births and deaths are given by 
$\tilde{\tau}_{0,k}=(\lambda_{0,k,1}, \lambda_{0,k,2}),k=1,2,\ldots,k_0$. 
The  computations in the steps that are summarized below are done using the R function \texttt{gridDiag} \citep{Fasy2014IntroductionTT}.

\begin{description} \label{TDAonfun}
\item [Step 1.] Assume a filtration parameter starting at $\lambda=\min_{\mathbf{z}} f(\mathbf{z})$ and let $L_\lambda(f) = \{\mathbf{z}: f(\mathbf{z})=\min_{\mathbf{z}} f(\mathbf{z})\}$. 

\item [Step 2.] 
Construct topological features for increasing values of $\lambda$. For each $\lambda$, simplicial complexes can be constructed from the sublevel set $L_\lambda(f)$, and $\tilde{\alpha}_{\tilde{p}, k}$ can be computed using computation homology, where $0\leq\tilde{p}\leq d-1$.
If an elder topology $\tilde{\alpha}_{\tilde{p},k_1}$ and a younger topology $\tilde{\alpha}_{\tilde{p},k_2}$ merge into a single  $\tilde{\alpha}_{\tilde{p},k}$ at some $\lambda$ value, then $\tilde{\alpha}_{\tilde{p},k_1}$ becomes $\tilde{\alpha}_{\tilde{p},k}$ and $\tilde{\alpha}_{\tilde{p},k_2}$ dies at  $\lambda$, using the \textit{Elder Rule} \citep{harer2010}.

\item [Step 3.] The persistence diagram is the output of the set of points representing birth-death of homology groups $\{\tilde{\tau}_{\tilde{p},k}=(\lambda_{\tilde{p},k,1}, \lambda_{\tilde{p},k,2}): \tilde{p}=0,1, \ldots;k=1,2,\ldots,k_{\tilde{p}}\}$.
\end{description}
\vskip1em

\begin{figure}
\centering
\includegraphics[width=\textwidth]{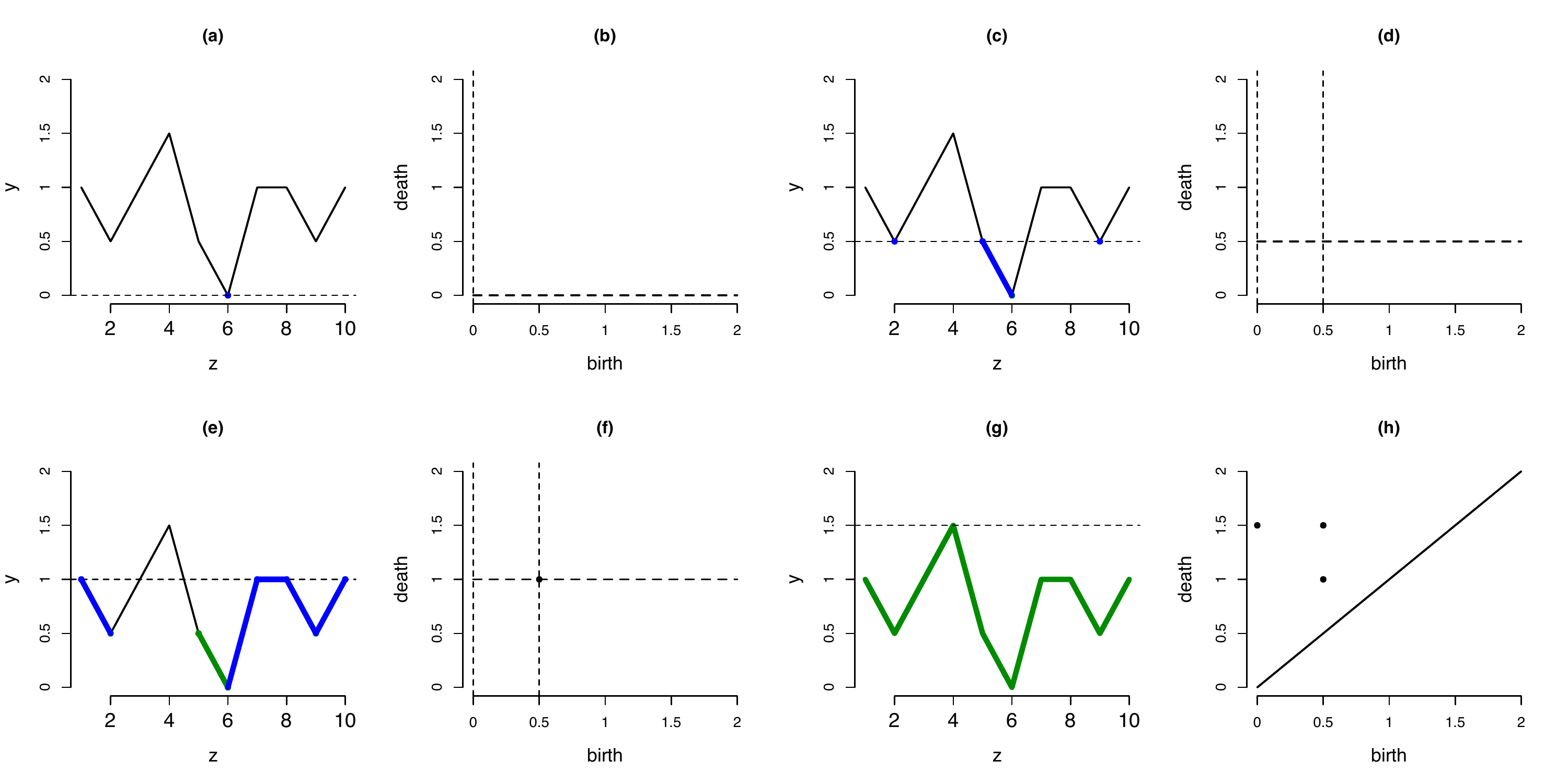}
\caption{Construction of a persistence diagram corresponding to a one-dimensional continuous real function. (a) is the  function and (h) is the persistence diagram. (c), (e) and (g) show the sublevel set filtration procedure, while (b), (d) and (f) are the intermediate steps for constructing the persistence diagram.}
\label{functionpersis}
\end{figure}

\noindent \textbf{Example 3.1. Discretized Function to Persistence Diagram.} 
We  present an example of using TDA on a one-dimensional discretized real function generated using the code chunk below, where \texttt{funval} contains values of the function taken over a grid:
\begin{verbatim}
funval = c(1, 0.5, 1, 1.5, 0.5, 0, 1, 1, 0.5, 1)
(pers.diag.4 <- gridDiag(FUNvalues = funval, sublevel = TRUE) )
$diagram
     dimension Birth Death
[1,]         0   0.0   1.5
[2,]         0   0.5   1.5
[3,]         0   0.5   1.0
\end{verbatim}
The persistence diagram contains three connected components born at $\lambda=0$ and $\lambda=0.5$, corresponding to the function having three local minima at these two distinct $\lambda$ values.    
In Figure \ref{functionpersis}(a), a connected component emerges when $\lambda = 0$ and is marked as a blue 
dot (it is the earliest/oldest connected component).

\begin{verbatim}
plot(funval, x=1:10, type = "l",yaxt='n',ylim = c(0,2), cex.axis=1.4, xlab = "z",
    ylab= "y", cex.lab= 1.3, cex=1.2, lwd=1.5, lty=1, pch=1, bty='n', main="(a)")
points(x=6, y=0, pch=16, col="blue", type = "p")
ticks<-c(0, 0.5, 1, 1.5, 2); axis(2,at=ticks,labels=ticks)
abline(a=0, b=0, lty=2, lwd=1,pch=1)
\end{verbatim}

The vertical dashed line in Figure \ref{functionpersis}(b) corresponds to the birth time $\lambda= 0$, while the horizontal dashed line tracks the current filtration parameter $\lambda$. There is no point on the birth/death plot yet, as no connected components have died at $\lambda=0$. 
\begin{verbatim}
plot(rep(0,11),x=0:10/5,type = "l",cex.axis=1.4,xaxt='n',xlab= "birth",yaxt='n',
ylim=c(0,2),ylab="death",cex.lab=1.3,cex=1.2,lwd=1.5,lty=2,pch=1,bty='n',main="(b)")
axis(1,at=ticks,labels=ticks); axis(2,at=ticks,labels=ticks); abline(v=0, lty=2)
\end{verbatim}

Figure \ref{functionpersis}(c) corresponds to $\lambda = 0.5$. There are two more connected components  
indicated by the blue dots at $(2,0.5)$ and $(9,0.5)$. The blue dot $(6,0)$ in the middle with a  blue 
line connecting it to the %dark green 
dot $(5,0.5)$ %on the left 
indicates that the oldest connected component enlarges. 
\begin{verbatim}
plot(funval, x=1:10, type = "l",yaxt='n',ylim = c(0,2),cex.axis=1.4,xlab = "z",
    ylab = "y",cex.lab= 1.3, cex=1.2,lwd=1.5, lty=1, pch=1, bty='n', main = "(c)")
axis(2,at=ticks,labels=ticks); abline(a=0.5, b=0, lty=2, lwd=1,pch=1)
points(x=6, y=0, pch=16, col="green4", type = "p",xlab = "z", ylab= "y")
points(x=c(2,5,9), y=rep(0.5,3), pch=16, col="blue", type = "p",xlab="z",ylab="y")
segments(x0=6, y0=0, x1=5, y1=0.5, lty = 1, pch=1, lwd=2.5, col = "blue")
\end{verbatim}

Figure \ref{functionpersis}(d) has one more vertical dashed line, which gives the birth time for the other two new connected components. There is no connected component dead yet, and so no points are shown on the second plot either.  The code chunks for plotting these are similar and are not shown due to space limitations.
When $\lambda = 1$, in Figure \ref{functionpersis}(e), all components enlarge and one newer component is killed by the elder one because they are merged. 
There is a black dot at $(0.5,1)$ in Figure \ref{functionpersis}(f), which indicates the newer connected component that is born at $\lambda=0.5$ and is dead at $\lambda=1$. 
When $\lambda = 1.5$ reaching the maximum of the function in Figure \ref{functionpersis}(g), the last component is killed.
The black dot at the location $(0,1.5)$ in Figure \ref{functionpersis}(h) is for the last component. The other black dots corresponding to $(0.5, 1.5)$ and $(0.5, 1)$ show the birth and death of other connected components.

\vskip1em

\noindent \textbf{Example 3.2. Morse Function to Persistence Diagram.} 
An alternate technique for persistent homology which is robust to noisy point cloud data uses the R function \texttt{gridDiag} \citep{JMLRChazal2018}. 
As mentioned in the introduction, a point cloud is assumed as a sample from an underlying manifold. 
To learn the topology of the manifold, \texttt{gridDiag} enables us to construct a Morse function such as the distance-to-measure (DTM) function from the point cloud using the sublevel set filtration.
Suppose the point cloud is $\{\mathbf{x}_i \in \mathcal{R}^d, i=1,2,\ldots,N\}$. Represent the DTM function %as $f^{(DTM)}: 
from $\mathcal{R}^d \rightarrow \mathcal{R}$ as
$$
f^{(DTM)}(\mathbf{x}) = \sqrt{\sum_{\mathbf{x}_i \in \mathcal{N}_k(\mathbf{x})} ||\mathbf{x}_i-\mathbf{x}||^2/k},
$$
where $k=[mN]$ ($m$ is the parameter \texttt{m0}) and $\mathcal{N}_k(\mathbf{x})$ is the set containing the $k$ nearest neighbors of $\mathbf{x}$ in the point cloud. 

A higher value of  $f^{(DTM)} (\mathbf{x})$ means that $\mathbf{x}$ is further away from most of the points. DTM is also robust to outliers \citep{JMLRChazal2018}. 
%since $f^{(DTM)}$ does not depend on all but majority of points in the point cloud, controling by its parameter \texttt{m0} (smaller \texttt{m0} would let DTM insensitive to a higher proportion of points). Therefore, it is more useful to persistent homology on a point cloud when the point cloud contains outliers deteriorating topological features, which are called critical points in the TDA literature. 
%
We illustrate using the same point cloud data from Example 2.1:

\begin{verbatim}
m0=0.05; by <- 0.065; Xlim <- range(PC[,1]); Ylim <- range(PC[,2])
(pers.diag.5 = gridDiag(X=PC, FUN=dtm, lim=cbind(Xlim, Ylim), by=by, m0=m0) )
$diagram
      dimension      Birth      Death
 [1,]         0 0.02414385 0.96912324
 [2,]         0 0.02623549 0.15885911
 [3,]         0 0.03489488 0.15662338
 ...
 [20,]         0 0.14527336 0.15092905
 [21,]         1 0.20234735 0.96912324
\end{verbatim}
Here, \texttt{lim} specifies the range of the point cloud in different dimensions, \texttt{by} is the step size for increasing $\lambda$, and \texttt{m0} (which lies in $(0,1)$ with 0.05 as the default value) is the smoothing parameter of the DTM method.  
The code for constructing Figure \ref{PCDTM} is shown below:

\begin{verbatim}
par(mfrow = c(1,2))
Xseq <- seq(from = Xlim[1], to = Xlim[2], by = by)
Yseq <- seq(from = Ylim[1], to = Ylim[2], by = by)
Grid <- expand.grid(Xseq, Yseq); DTM = dtm(X = PC, Grid = Grid, m0 = m0)
persp(x = Xseq, y = Yseq,z = matrix(DTM, nrow = length(Xseq), ncol = length(Yseq)),
      xlab = "", ylab = "", zlab = "", theta = -20, phi = 35, scale = FALSE,
      expand = 2, col = "red", border = NA, ltheta = 50, shade = 0.5,main = "(a)")
plot(pers.diag.5[["diagram"]], main = "(b)")
\end{verbatim}

\begin{figure}
\centering
\includegraphics[width=\textwidth]{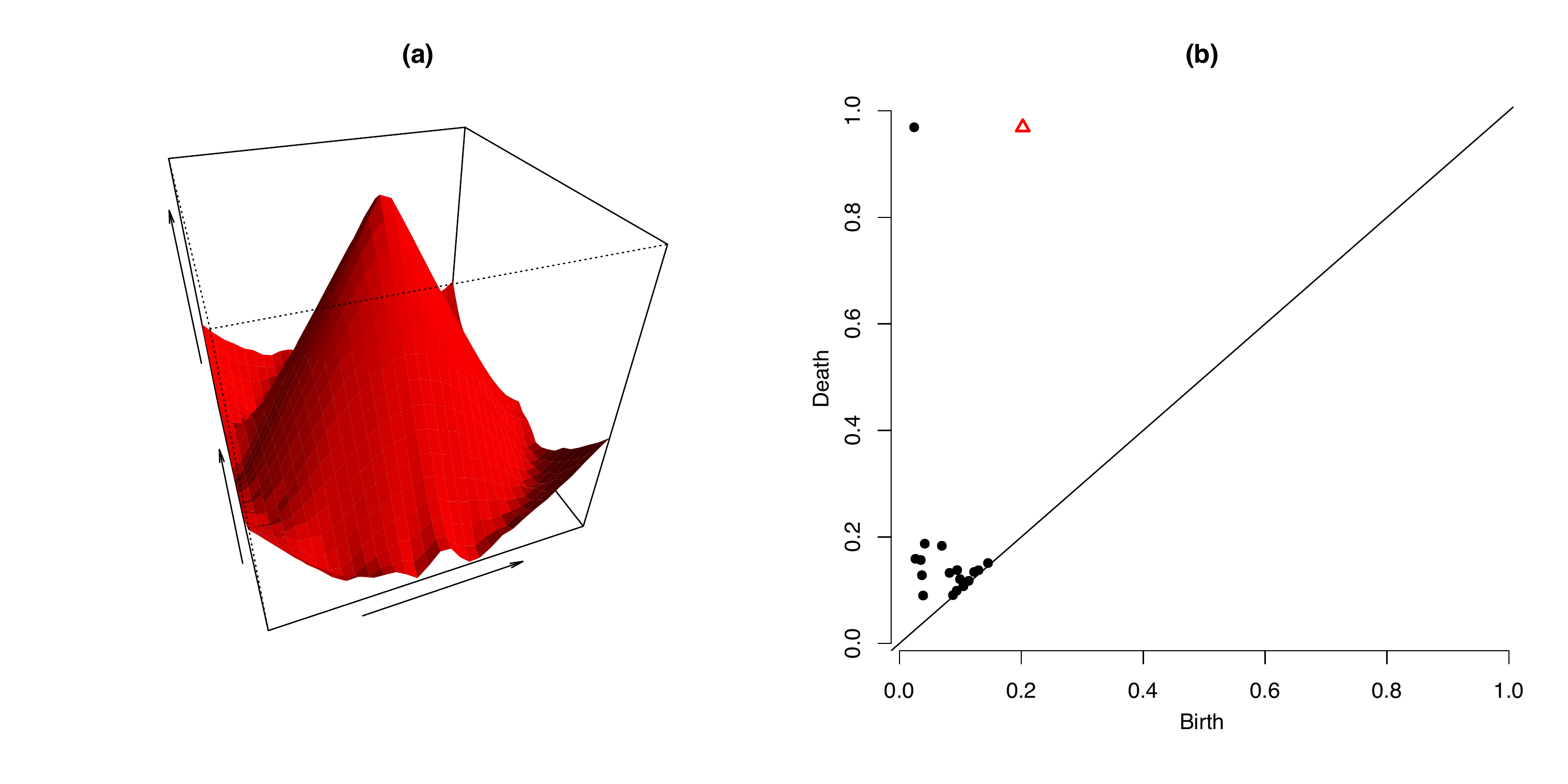}
\caption{Distance-to-Measure (DTM) function and the persistence diagram.}
\label{PCDTM}
\end{figure}

Figure \ref{PCDTM}(a) shows
the DTM function of the point cloud. 
The function has a 
peak in the middle of the plot,
surrounded by a rough circle of local minima (since the original point cloud is from the unit circle and DTM is a smoothed distance function)
Figure \ref{PCDTM}(b) shows the persistence diagram.  The dots in the persistence diagram are the birth-death points of $0$-th homology groups and the triangle denotes the birth-death point of $1$-th homology group. 
There are only 20 birth-death points of $0$-th homology groups instead of 60 as we saw in Example 2.1, because
the DTM smooths the distance function, as mentioned earlier,  resulting in a different output from those in Examples 2.1 and 2.2.
The point at $(0.2, 0.97)$ indicates a big circle in the data (representing the topology of the underlying unit circle).

\subsection{TDA of Time Series via Frequency Domain Functions} \label{tstdaspec}

Section 3.2.1 discusses TDA starting from variations of the Fourier transform for continuous-valued time series, while Section 3.2.2 shows how to build persistence diagrams based on Walsh Fourier transforms for categorical time series.

\subsubsection{Discrete Fourier Transforms to Persistence Diagrams}

In this section, we look at topological properties of time series through their frequency domain representations such as second-order spectra. 
We  construct a persistence diagram using  sublevel set filtration on  the smoothed  tapered estimate of the second-order spectrum of the time series $\{x_t, t=1,\ldots,T\}$.
The modified DFT \citep{Stoffer1991}  and corresponding periodogram with tapering are defined as 
\begin{eqnarray*}
d_{h}(\omega_j) &=& T^{-1/2} \sum_{t=1}^T h_t x_t e^{-2\pi i \omega_j t} \mbox{ and }\\
I_{h} (\omega_j) &=& |d_{h} (\omega_j)|^2,
\end{eqnarray*}
for $t=1,2,\ldots, T$, where $h_t$ is a taper function. 
In Example 3.3, we show the use of the R function \texttt{gridDiag} to construct the persistence diagram starting from a smoothed version of $I_{h} (\omega_j)$, using the Daniel window for smoothing. 

\vskip1em

\noindent \textbf{Example 3.3. Smoothed Tapered Second-order Spectrum to Persistence Diagram.}
We use the R function \texttt{gridDiag} to construct the persistence diagrams for the same three periodic time series signals shown in Example 2.4. 
For Case 1, \texttt{spc.t1} denotes the smoothed tapered periodogram (taper$=0.1$ as default in the R function \texttt{spec.pgram}, and smoothing via the modified Daniel window $(0.25, 0.5, 0.25)$) of the time series \texttt{ts1}. % The code for other cases is similar.

\begin{verbatim}
x.1 = 1:length(ts1); ts1 = lm(ts1~x.1)$residuals; ts1 = ts1/sd(ts1) #Step 1
spc.t1=spec.pgram(ts1, kernel=kernel("modified.daniell", c(1)), plot = F)#Step 2 
#Step 3
PD.t1=gridDiag(FUNvalues=spc.t1$spec,location = FALSE, sublevel=TRUE)$diagram 
\end{verbatim}

\begin{figure} [H]
\centering
\includegraphics[width=\textwidth]{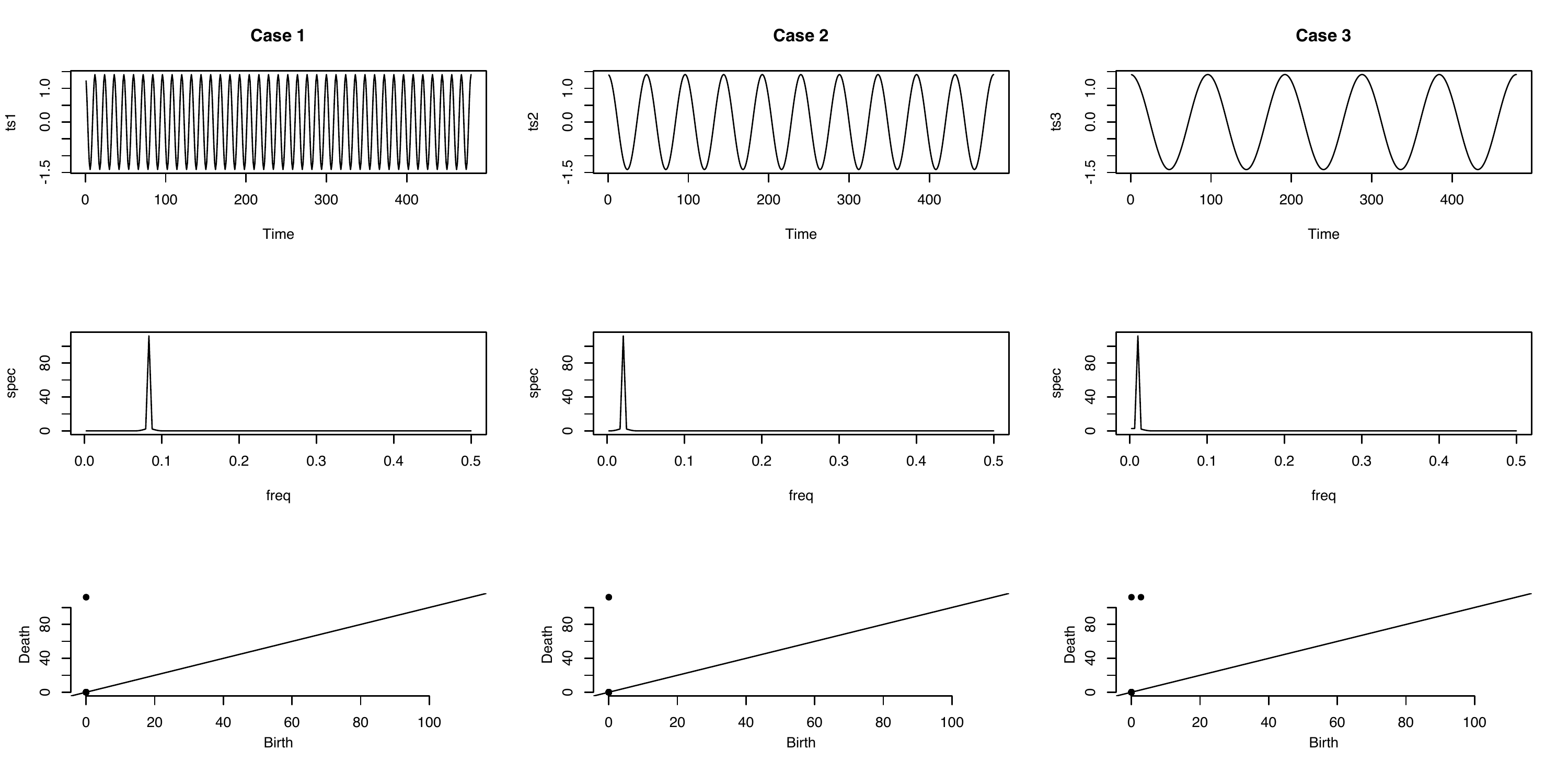}
\caption{Persistence diagrams using second-order spectrum.}
\label{specUsePCEx}
\end{figure}

In Figure \ref{specUsePCEx}, the top row shows the time series signals, the middle row shows the second-order spectra and the bottom row shows the persistence diagrams. The peaks in the spectra occur at different frequencies and correspond to the contributions at these frequencies to the total variance of $x_t$. 
The three persistence diagrams in the bottom row are similar to each other, since this method is insensitive to 
differences in the periodicity of the time series.

In contrast with the persistence diagrams in Figure 3 for the same signals, we no longer see red triangles for $1$-th homology groups, indicating that there is a difference between constructing persistence diagrams from point clouds versus the spectrum.

We compute the bottleneck distances between the three persistence diagrams. 
The distance between Case 1 and Case 2 is 0.01, which is much smaller than the distance between Case 1 and Case 3 which is 2.77, or the one between Case 2 and Case 3 which is 2.76. The code for computing the distances is similar to the one shown in Example 2.4.

\subsubsection{Walsh-Fourier Transforms to Persistence Diagrams}

\cite{Stoffer1991} suggested that Walsh spectral analysis is suited to  the analysis of discrete-valued and categorical-valued time series, and of time series that contain sharp discontinuities. 
The fast Walsh-Fourier Transform construction uses the method of \cite{Shanks1969} to decompose a time series 
$\{x_t, t=1,\ldots, T\}$  into a sequence of Walsh functions, each representing a distinctive binary sequency pattern. If the time series length $T$ is not a power of $2$, let $T_2$ denote the next power of $2$. For example, if $T=1440$, then  $T_2 = 2^{11} =2048$. 
We use zero-padding to obtain a time series of length $T_2$ by setting set  $x_{T+1}, x_{T+2}, \ldots, x_{T_2} = 0$. 

For $j=0,\ldots, T_2-1$, let $\lambda_j = j/T_{2}$ denote the $j$th sequency. 
Let $W(t, j)$ denote the $t$-th Walsh function value in sequency $\lambda_j$. Walsh functions are iteratively generated as follows \citep{Shanks1969}: 
\begin{eqnarray}
W(0,j) &=& 1, j=0, 1, \ldots, T_2-1, \notag  \\
W(1, j)&=& 
\begin{cases}
1 & j=0, 1,\ldots, (T_2)/2-1 \\
-1 &  j=(T_2)/2, (T_2)/2+1, \ldots, T_2-1
\end{cases}  \notag \\
W(t, j) &=& W([t/2],2j) \times W(t-2[t/2], j), \\
t&=&2, \ldots, T_2-1, \hspace{0.1in} j=0, 1, \ldots, T_2-1, \notag
\end{eqnarray}
where $[a]$ denotes the integer part of $a$. For more details on Walsh functions,  refer to \cite{Stoffer1991}.
The Walsh-Fourier Transform (WFT) of the time series is computed as
\begin{equation}
 d_T(\lambda_j) = \frac{1}{\sqrt{T_2}} \sum_{t=1}^{T_2} x_{t}  \ W(t, j)), \hspace{0.1in} 0\leq j \leq T_2 -1.
\end{equation}

The computational complexity is $O(T \log (T))$  \citep{Shanks1969}. 
In Example 3.4, %\ref{34WFTPD}
we illustrate the construction of a persistence diagram for a categorical time series with two levels.
\vskip1em

\begin{figure} [H]
\centering
\includegraphics[height=0.35\textheight,width=0.9\textwidth]{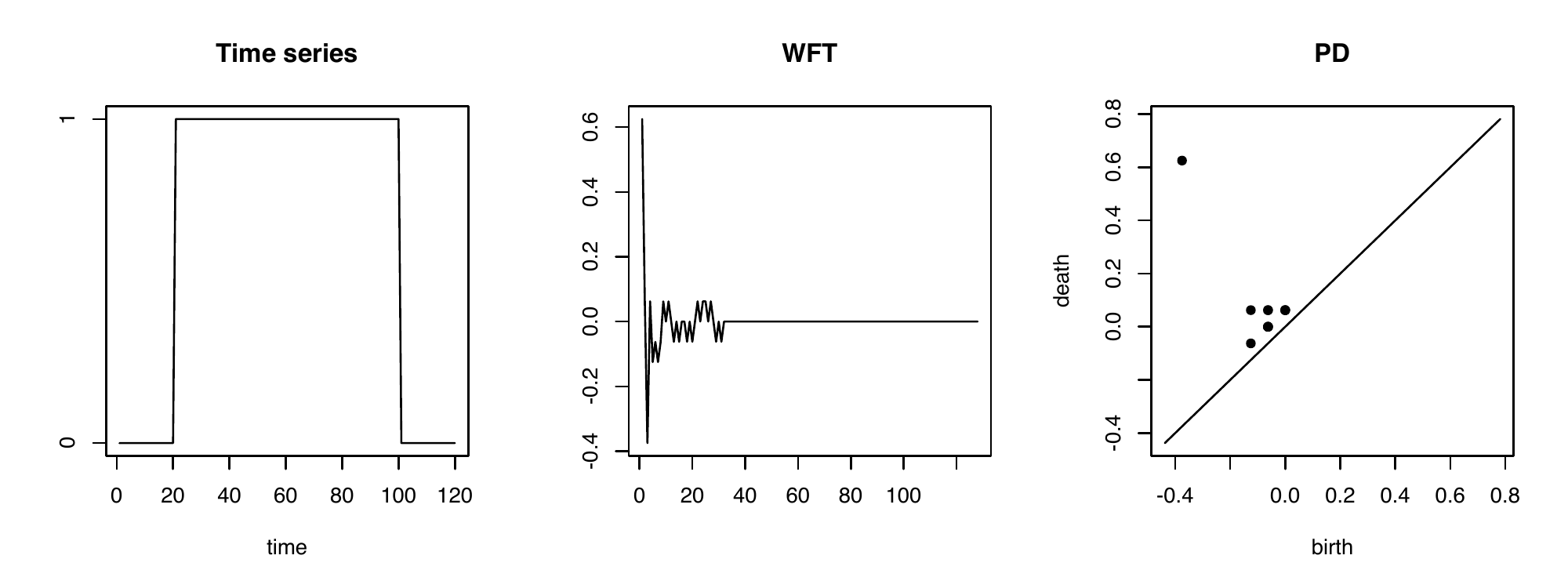}
\caption{Persistence diagrams using Walsh-Fourier Transforms.}
\label{surWFT}
\end{figure}

\noindent \textbf{Example 3.4. Walsh-Fourier Transform to Persistence Diagram.} %\label{34WFTPD}
In Figure \ref{surWFT}, a simulated categorical time series  of length $T=120$ with two levels, 0 or 1, is shown in the first column. Level 1 only occurs in the period between $t=21$ and  $t=100$. The middle column shows the WFT of the time series, while the third column shows its persistence diagram. There is one point in the diagram away from the diagonal line, which is a significant birth-death point of the $0$-th homology groups.

The R code for simulating the time series and converting the WFT into a persistence diagram is shown below. 
\begin{verbatim}
x.ts = c(rep(0, 20), rep(1, 80), rep(0, 20))
# create WFT using C++ code
x.diag=gridDiag(FUNvalues = x.WFTs, location = FALSE, sublevel = TRUE)$diagram
\end{verbatim}

\section{Feature Construction Using TDA}
\label{FeaturesTDA}

Unlike a vector space, the space of persistence diagrams is not easy to work with. For instance, a set of persistence diagrams may not have a unique mean \citep{Mileyko2011}. 
The bottleneck or Wasserstein distances are also more complicated than the Euclidean distance in practice. This section discusses an
alternative.

\subsection{Persistence Landscapes - A Basic Review}
\label{PersistenceL}

\cite{bubenik2015statistical} introduced persistence landscapes as useful statistical summaries which build topological features and are easy to combine with tools from statistics and machine learning.  
This section reviews persistence landscapes while Section \ref{PLTS} describes their construction for time series, with examples.

The $\nu$th order persistence landscape of  $\tilde{p}$-th homology groups is defined as
\begin{equation}
\mbox{\textit{PL}}_{\tilde{p}, \nu}(\ell) = \{\min(\ell-\lambda_{\tilde{p},k,1}, \lambda_{\tilde{p},k,2}-\ell)_+:  k=1,2,\ldots\}_{(\nu)} \label{PL1}
\end{equation}
where 
$\lambda_{\tilde{p},k,1}$ and $\lambda_{\tilde{p},k,2}$ were introduced under Step 3 of Section \ref{tda_pc}, 
$\ell \in \mathcal{R}$, 
$\min(a, b)_+$  denotes the smaller value if both $a$ and $b$ are positive, or zero if neither value is positive, 
%For example, $\min(-1, 1)_+=1$ and  $\min(-1,-2)_+=0$. 
and $\{A\}_{(\nu)}$ is the 
$\nu$-th order statistic of the set $A$. 
\cite{bubenik2015statistical} proved that a set of persistence landscapes admits a unique mean and preserves statistical stability of the data distribution.
If we assume 
that the observed data  $\mathbf{X}$ is a random draw from an underlying space $\mathcal{X}$ equipped with a probability measure, and assume multiple copies $\mathbf{X}_1, \mathbf{X}_2,\ldots, \mathbf{X}_N$, then the mean of their persistence landscapes $\overline{\textit{PL}}^{(N)}_{\tilde{p}, \nu}(\ell)=\sum_{i=1}^N \textit{PL}^{(i)}_{\tilde{p}, \nu}(\ell)/N$ would converge almost surely (as $N\rightarrow\infty$) to the expectation of the persistence landscapes $E(\textit{PL}^{\mathcal{X}}_{\tilde{p},\nu}(\ell))$
if and only if 
$$
E||\textit{PL}^{\mathcal{X}}_{\tilde{p},\nu}(\ell)|| = \int_\ell |\textit{PL}^{\mathcal{X}}_{\tilde{p},\nu}(\ell)|d\ell<\infty,
$$
i.e.,  $\int_\ell |\textit{PL}^{\mathcal{X}}_{\tilde{p},1}(\ell)|d\ell<\infty$ for all $\tilde{p}$. 
This means that the set of persistence landscapes from the observed data is a good representative for the underlying distribution of true persistence landscapes and it is possible to do statistical inference for these  using the sample persistence landscapes.    
Another important statistical property is the stability in terms of using persistence landscapes versus using persistence diagrams. Suppose there are two persistence diagrams $\tilde{\Omega}_1$ and $\tilde{\Omega}_2$ together with the corresponding persistence landscapes $\textit{PL}^{(1)}_{\tilde{p}, \nu}(\ell)$ and $\textit{PL}^{(2)}_{\tilde{p}, \nu}(\ell)$. Then,   $||\textit{PL}^{(1)}_{\tilde{p}, \nu}(\ell)-\textit{PL}^{(2)}_{\tilde{p}, \nu}(\ell)||_q= \big[\int_\nu|\textit{PL}^{(1)}_{\tilde{p}}-\textit{PL}^{(2)}_{\tilde{p}}|^q \big]^{1/q}$ is no larger than a function of the $q$-Wasserstein distance $\mathbf{W}_{q, \tilde{p}}(\tilde{\Omega}_1,\tilde{\Omega}_2)$. Intuitively, it means that using persistence landscapes could preserve differences in the persistence diagrams.

The following steps describe the construction of persistence landscapes of $\tilde{p}$-th homology groups  in increasing order starting from the first-order landscape.

\begin{description}
\item [Step 1.] Extract the $1$-th homology groups $\tilde{\tau}_{\tilde{p}, k}$ ($k=1,2,\ldots,k_{\tilde{p}}$) from the persistence diagram $\tilde{\Omega}$ and create a set of computation grids, $\ell=M_1, M_1+\delta, M_1+2\delta, \ldots, M_2$, where the lower and upper bounds $M_1$ and $M_2$ are usually set as $\min_k \lambda_{\tilde{p}, k,1}$ and $\max_k \lambda_{\tilde{p}, k,2}$ respectively. 
Here, $\delta$ depends on the degree of resolutions. %to proceed. 
The \texttt{R} function \texttt{landscape} uses 500 grid points by default so that  $\delta=(M_2-M_1)/500$.

\item [Step 2.] Use each $\tilde{\tau}_{\tilde{p}, k}$ to compute  $\textit{PL}_{\tilde{p},k}(\ell) = \min(\ell-\lambda_{\tilde{p},k,1}, \lambda_{\tilde{p},k,2}-\ell)_+$ for all values of $\ell$.

\item [Step 3.] Fixing $\ell$, sort $\textit{pl}_{\tilde{p},k}(\ell)$ in  decreasing order, calling them $\textit{pl}^{(1)}_{\tilde{p}}(\ell), \textit{pl}^{(2)}_{\tilde{p}}(\ell),\ldots, \textit{pl}^{(k_{\tilde{p}})}_{\tilde{p}}(\ell)$.

\item [Step 4.] Output the $\nu$th order persistence landscape of the $\tilde{p}$-th homology groups, $\textit{PL}_{\tilde{p},\nu}(\ell) =  (\textit{pl}^{(\nu)}_{\tilde{p}}(\ell))_+$, where $\nu=1,2,\ldots$ and set $\textit{pl}^{(\nu)}_{\tilde{p}}(\ell)=0$ when $\nu > k_{\tilde{p}}$.
\end{description}

Step 3 primarily determines the computational cost of constructing persistence landscapes. 
Higher order persistence landscapes require more values sorted %$\textit{pl}_{\tilde{p},k}(\ell)$ 
for each $\ell$, which could be costly when $k_{\tilde{p}}$ (number of $\tilde{p}$-th homology groups) is large. 
\vskip1em

\noindent \textbf{Example 4.1. Persistence Landscapes from Persistence Diagrams.}   % from Example 3.} 
The R function \texttt{landscape} can be used to compute persistence landscapes. We illustrate on the the persistence diagram from Example 2.3.

\begin{verbatim}
Diag = pers.diag.3$diagram; Land <- c(); k=1; threshold=1
while(threshold>0){
  Land <- cbind(Land, landscape(Diag = Diag, dimension = 0, KK = k,
            tseq = seq(min(Diag[,2:3]), max(Diag[,2:3]), length=500)))
  threshold = sum(abs(Land[, ncol(Land)]))        
}
            [,1] [,2] [,3] [,4]
[1,] 0.000000000    0    0    0
[2,] 0.003006012    0    0    0
[3,] 0.006012024    0    0    0
  .....
[499,] 0.003006012 0.003006012    0    0
[500,] 0.000000000 0.000000000    0    0
\end{verbatim}

In the code above, the function \texttt{landscape} takes several arguments. When  \texttt{dimension=0}, it takes $0$-th homology groups of the persistence diagram \texttt{pers.diag.3} to compute landscape functions. The \texttt{KK} argument specifies the order of landscapes to be computed. The \texttt{tseq} argument specifies the range of the landscape functions. It uses a while-loop to compute persistence landscapes of all orders (here, four) over $500$ grids ranging from the minimum of the birth time to the maximum of the death time. 

\begin{figure}[H]
\centering
\includegraphics[width=\textwidth]{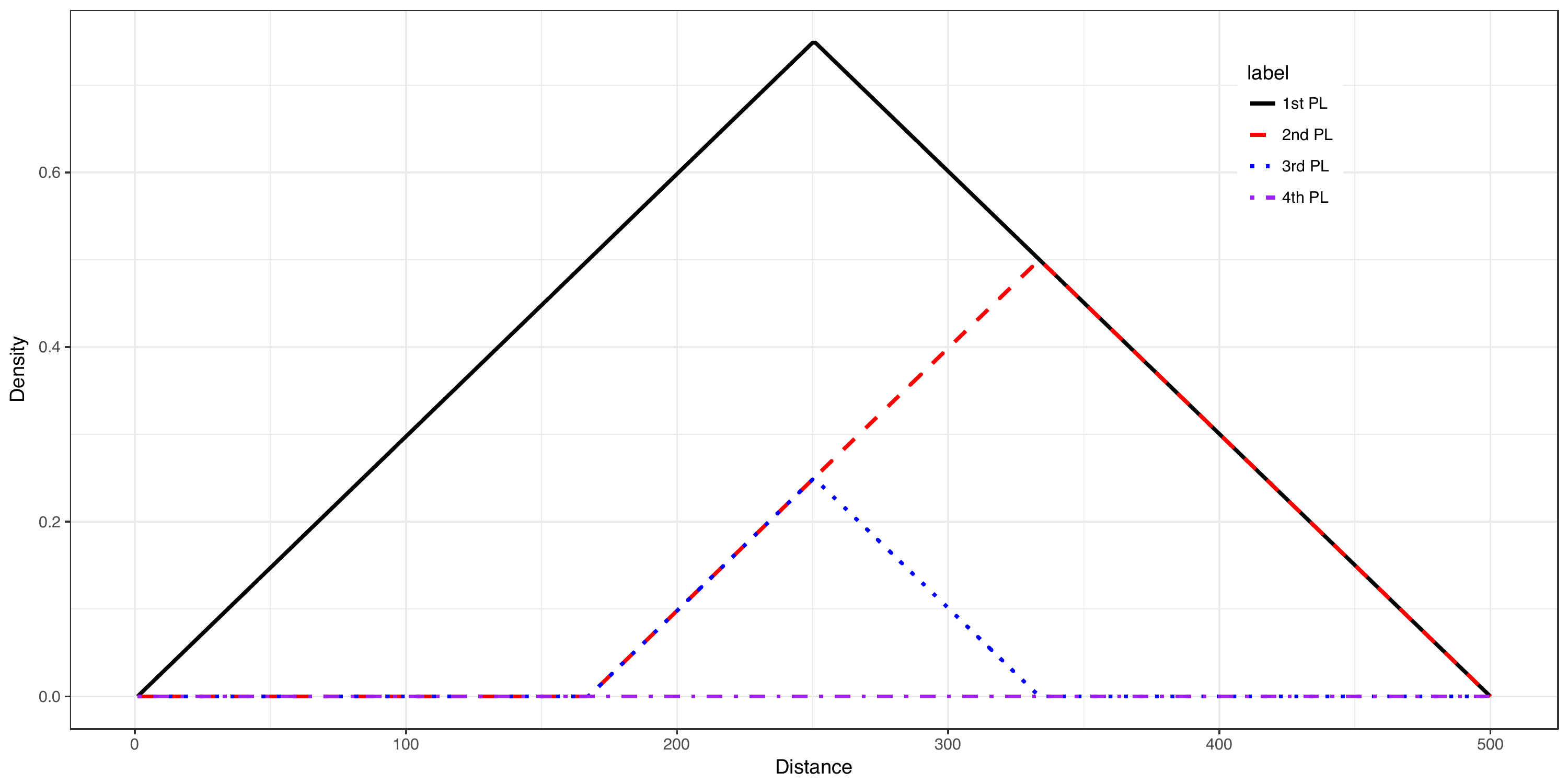}
\caption{Persistence landscapes of the persistence diagram \texttt{pers.diag.3}.}
\label{sec22funvalPL}
\end{figure}

Persistence landscapes of any order  
can be used as features.  
Lower order persistence landscapes contain information about important topological features than higher order persistence landscapes which are closer to zero and handle topological noise. Therefore, selecting the order of the persistence landscapes to serve as features requires a delicate balance between missing important signals and introducing too much noise.

\subsection{TDA of Time Series via Persistence Landscapes}\label{PLTS}

TDA on functions can be used to construct feature representations for time series analysis, and 
persistence landscapes 
are useful as topological representations for similarity/dissimilarity analysis on time series. In the literature, different %functional 
representations of time series have been used, such as the weighted Fourier transform in \cite{wang2018topological} or the Walsh-Fourier transform in \cite{chen2019clustering}. We describe these situations in the following sections.

\subsubsection{Persistence Landscapes for Continuous Time Series}\label{WangEEG}

\cite{wang2018topological} proposed TDA 
to measure structural changes in electroencephalogram (EEG) time series.
They first constructed Fourier transforms of the time series, then applied an exponential weighting scheme on the Fourier transforms to focus on the more important low frequency components of EEG. They further smoothed the weighted Fourier transform in order to make it a Morse function \citep{palais1963morse}.

The smoothed weighted Fourier series of a time series $\{x_t, t=1,\ldots,T\}$ has the form
$$
\hat{\mu}_{T_u}^k (t)=\sum_{j\in \mathbf{I}_1} e^{-(2j \pi /T)^2 \sigma} a_j \cos(2j\pi t/T) + \sum_{j\in \mathbf{I}_2} e^{-(2j\pi/T)^2\sigma}b_j \sin(2j\pi t/T),
$$
where $\mathbf{I}_1=\{j=0,1,2,\ldots, k: |a_j|>T_u\}$, $\mathbf{I}_2=\{j=1,2,\ldots, k: |b_j|>T_u\}$, $T_u=s \sqrt{2\log(n)}$, $a_j = \frac{2}{T}\sum_{t=1}^T x_t \cos(2j\pi t/T)$, $b_j = \frac{2}{T}\sum_{t=1}^T x_t \sin(2j\pi t/T)$ and $a_0=\sum_{t=1}^T x_t /T$. 
Here, $k$ is the degree deciding the highest frequency $[k/T]$ to be included in the representation (for $T=500$, they used $k=99$),  $n$ is the number of data points in each phase and $s$ is the median of
the absolute deviation (MAD) of the Fourier coefficients:

\begin{eqnarray}
a^{(m)} = \mbox{median}\{|a_i|, i=1,2,\ldots, k\}
\notag \\
b^{(m)} = \mbox{median}\{|b_i|, i=1,2,\ldots, k\}
\notag \\
s = \mbox{median}\{|a_i-a^{(m)}|, |b_j-b^{(m)}|, i,j=1,2,\ldots, k\}    
\end{eqnarray}

\iffalse
$$
s = \mbox{median}\{j=1,2,\ldots, k: |a_j-\mbox{median}\{i=1,2,\ldots, k: |a_i|\}|, |b_j-\mbox{median}\{i=1,2,\ldots, k: |b_i|\}|\}
$$
\fi
Using this finite sum of weighted sinusoidal functions, they argued that  $\hat{\mu}_{T_u}^k (t)$ becomes a Morse function. 
They used this Morse function representation to construct  persistence landscapes of all orders as features to detect possible structural changes. 
A main contribution of this paper is to show via simulation studies that the proposed TDA framework is robust to topology-preserving transformations such as translation and amplitude and frequency scaling, while being sensitive to topology-destroying transformations.
They argued the topological change happens only if there is a structural change in the time series.

\subsubsection{Persistence Landscapes for Categorical Time Series} \label{ChenTrans}

\cite{chen2019clustering} described TDA of categorical time series via their Walsh-Fourier transforms (which are not Morse functions). 
They constructed first order persistence landscapes based on Walsh-Fourier transforms of categorical time series, which they then used as features for  clustering. They applied this analysis to a large travel-activity data set, carrying out computations in parallel. 
They showed that construction of the first order persistence landscape only involves a linear transformation of the Walsh Fourier transform.

Given a sequence of WFT $d_T(n,\lambda_j), j=0, 1, \ldots, T_2-1$ of the time series $x_{n,t}, n=1,2,\ldots, N$, 
denote the minimum and maximum of the WFT values of the time series $x_{n,t}$ by 
$$
d_{n, \min}   = \min_{j} d_T (n,\lambda_j)  \mbox{ and }  d_{n, \max}= \max_{j} d_T (n,\lambda_j).
$$
Let 
$$
D_{\min} =  \min_{n} d_{n, \min}  \mbox{ and } D_{\max} = \max_{n} d_{n, \max}
$$ 
denote the minimum and maximum values of the WFTs across all $N$ time series.
The first-order persistence landscape of $x_{n,t}$ is obtained for $\ell = 1, 2,\ldots, L$ as  
\begin{equation}
\mbox{\textit{PL}}(n,\ell) = \min (V_1(n,\ell), V_2(n,\ell))_{+}
\label{PL1}
\end{equation}
where
\begin{eqnarray*}
V_1(n,\ell) = D_{\min} + \frac{(\ell-1) (D_{\max}-D_{\min})}{L-1} - d_{n, \min}, \nonumber\\
V_2(n,\ell) = d_{n,\max} - D_{\min} - \frac{(\ell-1) (D_{\max} - D_{\min})}{L-1}, 
\end{eqnarray*}
and $(a)_+$ denotes the positive part of a real number $a$. %Note that $\mbox{PL}(n, l) $ is a piecewise linear function.
For $\ell = 1, 2,\ldots, L$ and $n=1,\ldots,N$, the $\mbox{\textit{PL}}(n, \ell)$ are piecewise linear functions that constitute features constructed for each of the $N$ time series and useful for clustering. Our  \texttt{C++} code is available here: \url{https://github.com/bluemarlon/TDA-of-K-means-on-1st-PL}. %\UrlFont{https://github.com/bluemarlon/TDA-of-K-means-on-1st-PL}.

\begin{figure} [H]
\centering
\includegraphics[height=0.45\textheight, width=\textwidth]{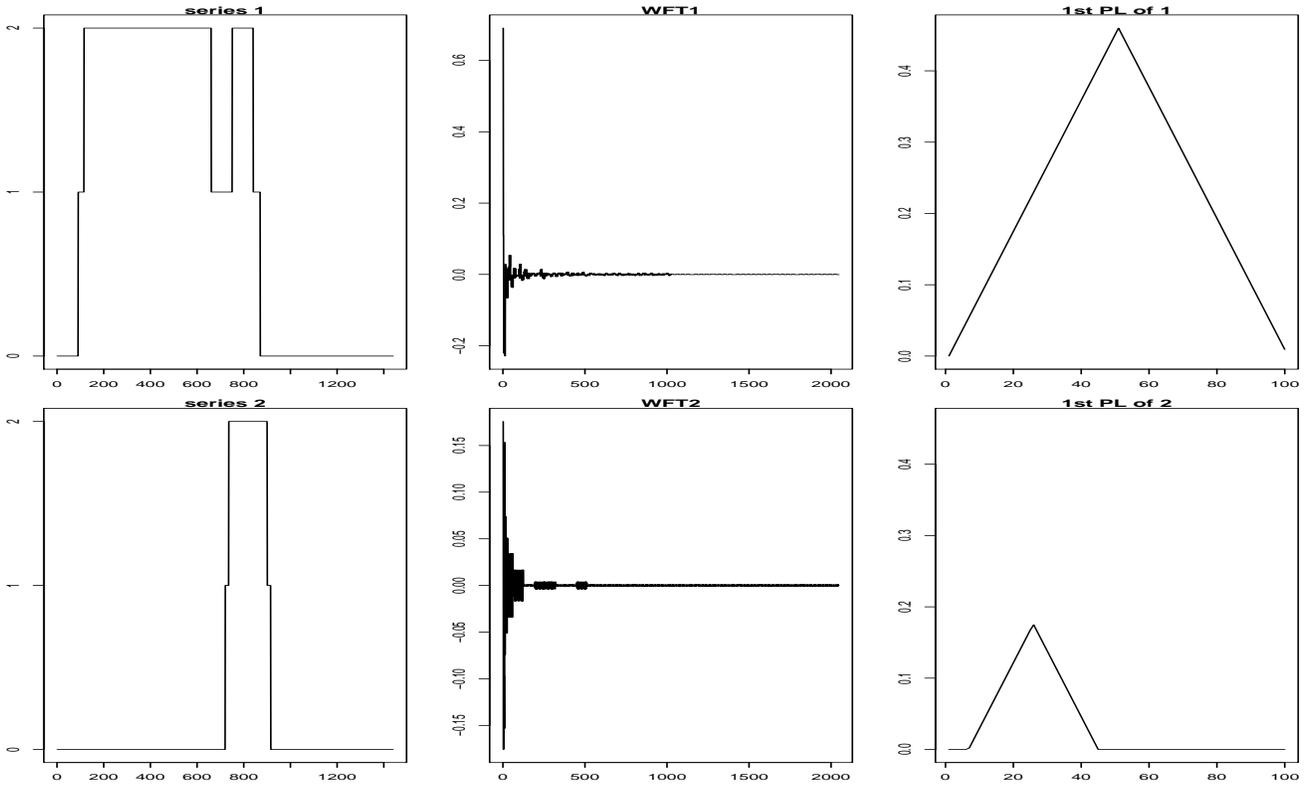}
\caption{Persistence diagrams using Walsh-Fourier Transforms.}
\label{WFTCatePD3}
\end{figure}

In Figure \ref{WFTCatePD3}, the first column shows categorical time series on activity-travel behavior of two randomly chosen adults from the National Household Travel Survey \citep{chen2019clustering}. The length of each time series is $T=1440$, corresponding to the number of minutes in a day. The response has three levels for each adult: 0 for staying at home, 1 for travel and 2 for being out of the home. 
The middle column shows the WFT of the time series, each of length $T_2=2048$. 
The last column shows first order persistence landscapes which are quite distinct for the two  series.

% or{red}{RC: you need a para to summarize the divide and combine K-means approach ypu used and the take-away. Just like ytou had a para for the previous section about Eang eo al.}

After the first order persistence landscapes are constructed, \cite{chen2019clustering} used a divide and combine K-means approach for clustering  a large number of subjects and 
%with three clusters to 
identified three distinct temporal patterns among them. The main contribution of this paper was to implement clustering of a large set of  activity-travel time series  through TDA of non-Morse functions. 
%for time series analysis. 
%Besides, the paper showed that there is a linear function to transform a univariate function into the first order persistence landscape, which is equalivant to using sublevel set filtration on the unvariate function and then converting the persistence diagram into the first order persistence landscape. 

\subsection{Other TDA Based Approaches}\label{OthersFeatures}

Numerical summaries other than persistence landscapes have been used  for  clustering \citep{berwald2013automatic,Pereira2015,Seversky2016}, classification \citep{YuheiUmeda2017D}, and break detection \citep{gidea2017topological,gidea2018topological} of time series. We give a brief review  in the following sections.

\subsubsection{Clustering} \label{TDAonClustering}

TDA based feature construction may be used with classical clustering methods, like K-means clustering for time series. 
For instance, \cite{Pereira2015}  
considered the model analyzed by \cite{costantino1995experimentally} for evaluating two regimes of adult Tribolium flour population growth (numerical measures) under stable equilibrium and aperiodic oscillations. They simulated
a total of 400 time series from the model for a period of 240 weeks (2 weeks per unit), consisting of 200 time series for the stable equilibrium regime and the other 200 for the aperiodic oscillations.
A brief description of their approach follows. 

\begin{description}
\item [Step 1.] They pre-processed the data using the Empirical Mode Decomposition (EMD) \citep{huang1998empirical} to denoise the data.

\item [Step 2.] They constructed point clouds using Takens' embedding with $d=2$ and $\tau=3$, taking the 
time series $\{x_t, t=1,2,\ldots,T\}$ to  $\{\mathbf{v}_i=(x_i, x_{i+3})| i=1,2,\ldots, T-3\}$. 
The point cloud, therefore, contains a total of $N=T-3$ number of points. 

\item [Step 3.] They used the  Witness complex \citep{de2004topological} for doing persistent homology.
Witness complex is essentially a different simplicial complex, which is most useful when dealing with large data sets. 
\item [Step 4.] After acquiring the persistence diagram $\tilde{\Omega}=\{\tilde{\tau}_{\tilde{p}, k}| k=1,2,\ldots,k_{\tilde{p}},\quad \tilde{p}=0,1\}$ ($\tilde{\tau}_{\tilde{p}, k} = (\lambda_{\tilde{p}, k,1}, \lambda_{\tilde{p}, k,2})$), they constructed the following set of features:

\begin{itemize}
\item[(i)] The number of points on each dimension $\tilde{p}$, i.e., $(k_{0}, k_1, k_2,\ldots, k_{\tilde{p}})$;

\item[(ii)] The maximum lifetime of each $\tilde{p}$, i.e,  $\max_{k}(\lambda_{\tilde{p}, k,2}-\lambda_{\tilde{p}, k,1})$;

\item[(iii)] The number of relevant points for each $\tilde{p}$, namely $\#\{\tilde{\tau}_{\tilde{p}, k}| \lambda_{\tilde{p},k,2}-\lambda_{\tilde{p},k,1}\geq 0.5 \max_k(\lambda_{\tilde{p},k,2}-\lambda_{\tilde{p},k,1})\}$, where $\#$ denotes the cardinality of the set;

\item[(iv)] The average lifetime of all homology groups on each $\tilde{p}$, $\sum_{k=1}^{k_{\tilde{p}}} (\lambda_{\tilde{p}, k, 2}-\lambda_{\tilde{p}, k, 1})/k_{\tilde{p}}$;

\item[(v)] The sum of the lifetimes of all homology groups on each $\tilde{p}$, $\sum_{k=1}^{k_{\tilde{p}}} (\lambda_{\tilde{p}, k, 2}-\lambda_{\tilde{p}, k, 1})$.
\end{itemize}
\item [Step 5.] Apply all features from Step 4 to do K-means clustering.
%with the number of clusters set at $k=2$. 
\end{description}

The implementation was done in \texttt{Java}. 
They compared  results with true labels using different metrics from the confusion matrix, and  %(i.e. recall, precision, F1-score, etc). 
showed that their method achieved a high F1-score of  $94\%$. 
In sum, they used EMD to filter out data noise, constructed point cloud using Takens' embedding method and used persistent homology to construct features for unsupervised learning. 
This logic is applicable to all other research papers. 

\subsubsection{Classification} \label{ClassifcationTS}

Features constructed using TDA  can be applied for classification of time series as well. 
\cite{YuheiUmeda2017D} described an example with motion sensor data of daily and sports activities used in \cite{altun2010human,altun2010comparative}, including both chaotic and non-chaotic time series data. 
%A total of 19 different activities were  treated as different clusters. 
Each of 19 activities was performed by 8 subjects and 60 signals were obtained for each activity and each subject, yielding  
%$19\times 8\times 60=
$9120$ time series. The sensor frequency was 512Hz and each signal was collected for 5 minutes.
%, resulting in a total of 
%$512\times 60\times 5=
%$153600$ time points for each time series. 
The pipeline of their method which was implemented in \texttt{MATLAB} is shown below.

\begin{description}
\item [Step 1.] Construct a point cloud via  Takens' embedding, $\{x_t, t=1,2,\ldots,T\} \rightarrow \{\mathbf{v}_i\} \subset \mathcal{R}^3,$ for $i=1,2,\ldots, T-2$ and $\mathbf{v}_i = (x_i, x_{i+1}, x_{i+2})$.

\item [Step 2.] Convert the point cloud to the persistence diagram $\tilde{\Omega}=\{\tilde{\tau}_{\tilde{p},k}| k=1,2,\ldots, \tilde{p}, \quad \tilde{p}=0,1,2\}$ using Rips complex and persistent homology. 

\item [Step 3.] Compute features using Betti sequences of $\tilde{p}$-th homology groups $\tilde{\Delta}_{\tilde{p}}(\lambda) = \#\{\lambda_{\tilde{p},k,1}\leq\lambda\leq \lambda_{\tilde{p},k,2}| k=1,2,\ldots, k_{\tilde{p}}\}$ from the persistence diagram,  
discretized into $300$ points for each $\tilde{p}$, and  connected into one feature vector of length $900$.

\item [Step 4.] Apply a one-dimensional convolutional neural network (CNN) \citep{krizhevsky2012imagenet} on the feature vector from Step 3 to do the classification. 
\end{description}

They concluded that their approach performed better than an approach using support vector machine (SVM) \citep{hastie01statisticallearning}.

\subsubsection{Structural Break Detection} \label{changepointTS}

TDA based features for structural break detection has been discussed in bioinformatics \citep{wang2018topological}, financial data analysis \citep{gidea2017topological,truong2017exploration,gidea2018topological,gidea2018topologicalcry}, etc. Here we describe the  approach in \cite{gidea2018topologicalcry}, who 
used four major daily log-price cryptocurrencies (Bitcoin, Ethereum, Litecoin, and Ripple) between 2016-01-01 to 2018-02-28 independently for break detection of critical transitions:

\begin{description}
\item [Step 1.] They constructed point clouds each with $50$ points from each log-price time series  $x_{t}, t=1,2,\ldots, T$. Each of the $T-52=448$ point clouds had a total of 50 points, the first point cloud being 
%say P^1 = 
$(v_1, v_2, \ldots, v_50)$ and the last point cloud being $(v_448, v_449, \ldots, v_497)$, with 
%and 
%\rightarrow \{
$\mathbf{v}_i = (x_{i}, x_{i+1}, x_{i+2}, x_{i+3})\in \mathcal{R}^4\}$. 

%These points are used for time-varying point clouds with size $50$.  or{blue}{Each windowing point cloud $\mathcal{P}^{(w)}=\{\mathbf{v}_w, \mathbf{v}_{w+1}, \ldots, \mathbf{v}_{w+49}\}$ has a total of 50 data points. There is a total of $T-52$ point clouds, $w=1,2,\ldots, T-52$.} 

\item [Step 2.] On each windowing point cloud $\mathcal{P}^{(w)}$, they constructed
%persistent homology using Rips complex is used and 
a persistence diagram $\tilde{\Omega}^{(w)}$ using Rips complex. Since they only focused on the $1$-th homology groups, $\tilde{\Omega}^{(w)}=\{\tilde{\tau}^{(w)}_{1, k},k=1,2,\ldots, k_1\}$.

\item [Step 3.] They constructed all persistence landscapes $\textit{PL}^{(w)}_{ 1, \nu}(\ell)$ from each $\tilde{\Omega}^{(w)}$,
$\nu=1,2,\ldots$, and then converted the persistence landscapes $\textit{PL}^{(w)}_{ 1, \nu}(\ell)$ to its L$_1$ norm, which is defined as $|| \textit{PL}^{(w)}_{1} ||_1 = \sum_{\nu=1}^\infty ||\textit{PL}^{(w)}_{ 1, \nu} ||_1$, where $||\textit{PL}^{(w)}_{ 1, \nu} ||_1=\int_{\mathcal{R}}\textit{PL}^{(w)}_{ 1, \nu}(\ell) d\ell$.

\item [Step 4.] On each window, they combined the log-price time series  $x_w$, first difference of the log-price values $x_{w+1}-x_{w}$ and the L$_1$ norm $|| \textit{PL}^{(w)}_{1} ||_1$ as the feature vector $(x_w, x_{w+1}-x_{w}, || \textit{PL}^{(w)}_{1} ||_1)$ and used it to do K-means clustering with number of cluster $K=18$.
\end{description}

They applied the method independently to each daily log-price cryptocurrency time series, and then used the clusters to identify topologically distinct regimes before the crash of each asset. They argued that this method has the potential to automatically recognize approaching critical transitions in the cryptocurrency markets, even when the relevant time series exhibit a highly non-stationary, erratic behavior.

\section{Discussion and Summary} \label{summary}

This paper gives a comprehensive overview of TDA which consists of a set of powerful tools for measuring topological features of time series and using it for pattern detection, clustering, classification, and structural break detection. Research extensions in several directions are possible.
First, TDA for 
multivariate time series analysis has been studied very recently 
\citep{gidea2018topological,gidea2017topological,stolz2017persistent}. 
A second extension consists of using summary statistics and dissimilarity measures for TDA. A review of summary statistics and dissimilarity measures refer to \cite{Nanopoulos2001,Fulcher2017FeaturebasedTA,Aghabozorgi2015}. 
Third, 
research into improved computation tools in computational topology and further exploration of statistical properties while using TDA is a rich research area. Ongoing research can be separated into these different scenarios: computational homology \citep{phillips2013geometric,de2004topological,liu2012fast,carlsson2009theory, 2018arXiv180910231C,ren2018weighted}, study of topological summaries \citep{bendich2016persistent, Christophe2019, Reininghaus2015ASM, kusano2016persistence, Carriere2017}, and statistical inference \citep{fasy2014, phillips2013geometric, JMLRChazal2018, ALAA2017406,robinson2017hypothesis,brecheteau2019}.

% Submissions are not required to reflect the precise reference formatting of the journal (use of italics, bold etc.), however it is important that all key elements of each reference are included.
%\bibliographystyle{jasa}
%\bibliography{thesis}

\end{document}